\documentclass[a4paper,10pt]{article}
\usepackage[utf8]{inputenc}
\usepackage[T1]{fontenc}
\usepackage{indentfirst}

\usepackage{amsmath,amssymb,amsthm,bbm}
\usepackage{enumerate} 
\usepackage[toc,page]{appendix}
\usepackage{authblk}
\usepackage[round,sort,comma]{natbib}
\usepackage[svgnames]{xcolor}
\usepackage{graphicx}
\usepackage{epstopdf}
\usepackage{booktabs}
\usepackage{multirow}

\usepackage{lipsum}

\newcommand\blfootnote[1]{%
  \begingroup
  \renewcommand\thefootnote{}\footnote{#1}%
  \addtocounter{footnote}{-1}%
  \endgroup
}

\newtheorem{theorem}{Theorem}[section]

\newtheorem{corollary}[theorem]{Corollary}

\newtheorem{lemma}[theorem]{Lemma}

\newtheorem{remark}[theorem]{Remark}

\title{Higher order approximation of call option prices under stochastic volatility models.}
\author[1]{Archil Gulisashvili}
\author[2,3]{Ra\'{u}l Merino}
\author[4]{Marc Lagunas}
\author[2]{Josep Vives}
\affil[1]{Department of Mathematics, Ohio University, Athens OH 45701.  }
\affil[2]{Facultat de Matem\`{a}tiques, Universitat de Barcelona, Gran Via 585, 08007 Barcelona, Spain}
\affil[3]{VidaCaixa S.A., Investment Risk Management Department, C/Juan Gris, 2-8, 08014 Barcelona, Spain.}
\affil[4]{Department of Mathematics, University of Oslo, P.O. Box 1053 Blindern, 0316 Oslo, Norway. }

\date{\today}
\begin{document}

\maketitle

\begin{abstract}
In the present paper, a decomposition formula for the call price due to Al\`{o}s is transformed into a Taylor type formula containing an infinite series with stochastic terms. The new decomposition may be considered as an alternative to the decomposition of the call price found in a recent paper of Al\`{o}s, Gatheral and Radoi\v{c}i\'{c}. We use the new decomposition to obtain various approximations to the call price in the Heston model with sharper estimates of the error term than in the previously known approximations. One of the formulas obtained in the present paper has five significant terms and an error estimate of the form $O(\nu^{3}(\left|\rho\right|+\nu))$, where $\nu$ is the vol-vol 
parameter, and $\rho$ is the correlation coefficient between the price and the volatility in the Heston model. Another approximation formula contains seven more terms and the error estimate is of the form
$O(\nu^4(1+|\rho|)$. For the uncorrelated Hestom model ($\rho=0$), we obtain a formula with four significant terms and an error estimate $O(\nu^6)$. Numerical experiments show that the new approximations to the call price perform especially well in the high volatility mode.   
\end{abstract}

\blfootnote{The authors would like to thank Jan~Posp\'{\i}\v{s}il for helpful discussions.}

\section{Introduction}
Stochastic volatility models were introduced in order to account for various downsides in the constant volatility assumption, on which the celebrated Black-Scholes model is based. One of the most popular stochastic volatility models is the Heston model, developed in \cite{Heston}. For more information on stochastic volatility models, see, e.g., \cite{Ga}. 

In the present paper we derive sharp approximation formulas for the call price in the Heston model. These high order approximations improve the previously known ones. In \cite{A06} and \cite{A12}, special decompositions of the call option price in the Heston model were found, using Malliavin calculus and Itô calculus, respectively. The main difference between those two decompositions is that the former one uses the average of future variances, while the latter one is based on the conditional expectation of such average. Note that the stochastic process consisting of the conditional expectation of future variances is an adapted process, whereas the process of the genuine average is an anticipating process. In \cite{A12}, an approximation formula with a general error term was obtained for the call price in the Heston model. This error term was quantified in \cite{ADV}, where it was shown that the error term has the form $O(\nu^{2}\left(\left|\rho\right|+\nu\right)^{2})$. In the previous expression, $\nu$ is the vol-vol parameter and $\rho$ is the correlation coefficient in the Heston model. However, in the above-mentioned approximation formula, some terms of order $\nu^{2}$ were ignored, whereas other terms of the same order were kept. This may be considered as a drawback in the approximation formula obtained in \cite{ADV}. Among the other earlier works, we would like to mention the paper \cite{MV15}, where the expansion obtained in \cite{A12} was extended to general stochastic volatility models of diffusion type. Moreover, in \cite{MerinoPospisilSobotkaVives17}, a general decomposition formula for a smooth functional of the log-price process was obtained for a general stochastic volatility model, as well as a decomposition formula for call options in models with finite activity jumps in the spot. Theorem 3.1 in \cite{MerinoPospisilSobotkaVives17} is used recursively in the present paper to approximate the exact call price decomposition obtained in \cite{A12} by an infinite series of stochastic terms. The first two terms in the new expansion are the same as in \cite{A12} and \cite{ADV}. Moreover, our result is consistent with the one obtained in \cite{AGR}, but presented and obtained in a different way. Using the new general approximation formula in the case of the Heston model, we add two more significant terms to the above-mentioned expansion, to reach an error of the form 
$O(\nu^{3}(\left|\rho\right|+\nu))$ (see Theorem \ref{BS 2nd Approximation}), and seven more significant terms to obtain an error estimate 
$O(\nu^4(1+|\rho|)$ (see Theorem \ref{BS Approximation2}). In the particular case of zero correlation, we derive an approximation formula with four terms with an error of order $O(\nu^6).$

We will next  briefly describe the structure of the paper. In Section \ref{sec:preliminaries}, we provide preliminary information and discuss the notation used in the paper. In Section \ref{sec:decomposition}, we establish a general decomposition formula, and show how to use it recursively to obtain higher-order approximation formulas for the call price. In Section \ref{sec:Heston}, we obtain two new approximation formulas for the call price in the Heston model (see Theorems \ref{BS 2nd Approximation} and \ref{BS Approximation2}). The error estimates in those formulas are of the order $O(\nu^{3}(\left|\rho\right|+\nu))$ and $O(\nu^4(1+|\rho|))$, respectively. We derive also an approximation formula of order $O(\nu^6)$ for the zero correlation case (see Corollary \ref{BS Approximation zero-corr}). In Subsection \ref{SS:numa} we  provide and discuss some numerical results. Final conclusions can be found in Section \ref{sec:conclusion}.

\section{Preliminaries and Notations.}\label{sec:preliminaries}

Let $T> 0$ be the time horizon, and let $W$ and $\widetilde{W}$ be two independent Brownian motions defined on a complete probability space $\left(\Omega ,\mathcal{F}, P\right).$ Denote by 
$\mathcal{F}^{W}$ and $\mathcal{F}^{\widetilde{W}}$ the filtrations generated by $W$ and $\widetilde{W}$, respectively. Set $\mathcal{F}_t:=\mathcal{F}^{W}_t\vee \mathcal{F}^{\tilde{W}}_t$, $t\in[0,T].$

Consider a stochastic volatility model in which the asset price process $S=\{S_{t}, t\in [0,T]\}$ satisfies the stochastic differential equation

\begin{equation}\label{stock}
dS_{t}=rS_{t}dt+\sigma _{t}S_{t}\left( \rho dW_{t}+\sqrt{1-\rho ^{2}}d\tilde{W}_{t}\right),   
\end{equation}
where $r\geq 0$ is the interest rate and $\rho \in (-1,1)$. The volatility process $\sigma$ is a square-integrable process adapted to the filtration generated by $W$, and it is assumed that the paths of the process $\sigma$ are strictly positive $P$-a.s. It is also assumed that $P$ is a risk-free measure, that is, the discounted asset price process $t\mapsto e^{-rt}S_t$, $t\in[0,T]$, is a martingale. The initial condition for the process $S$ will be denoted by $s_0> 0$.

We will mostly work with the log-price process $X_{t}=\log S_{t}$, $t\in[0,T]$. It satisfies the equation

\begin{eqnarray}\label{log-price model}
dX_{t}= \left(r - \frac{1}{2}\sigma^{2}_{t}\right)dt + \sigma_{t} \left(\rho dW_{t}  + \sqrt{1-\rho^{2}}d\tilde{W}_{t} \right),
\end{eqnarray}
and the initial condition is given by $x_0=\log s_0$.

The following notation will be used throughout the paper:

\begin{itemize}
\item 
$E_t:=E(\cdot|\mathcal{F}_t).$

\item 
The Black-Scholes function will be denoted by $(BS)$. It is given by

\begin{eqnarray}
\nonumber (BS)\left(t,x,y\right)= e^{x} \Phi(d_{+}) - K e^{-r\tau} \Phi(d_{-}),
\end{eqnarray}
where $\tau=T-t$ is the time to maturity, $y$ is the constant volatility, $K$ is the strike price, $r$ is the interest rate, and 
$\Phi$ denotes the cumulative distribution function of the standard normal law. The symbols $d_{+}$ and $d_{-}$ stand for the following functions:

\begin{eqnarray}
\nonumber d_{\pm} = \frac{x-\ln K + (r \pm \frac{y^{2}}{2})\tau}{y\sqrt{\tau}}.
\end{eqnarray}

\item 
The call option price in the Black-Scholes model is given by 

$$V_{t}=e^{-r(T-t)}{E}_t [(e^{X_{T}}-K)^+].$$

\item  
It is known that the function $(BS)$ satisfies the Black-Scholes equation ${\mathcal L}_{y}(BS)(t,x,y)=0,$ where

\begin{eqnarray}\label{FK}
\mathcal{L}_{y}:= {\partial}_t + \frac{1}{2}y_{t}^{2} {\partial^{2}_{x}} + \left(r -\frac{1}{2}y^{2}_{t}\right) {\partial}_{x}-r.
\end{eqnarray}

\item 
The following differential operators will be used in the paper: $\Lambda:=\partial_x$, $\Gamma:=\left(\partial^{2}_{x}-\partial_{x}\right)$, and $\Gamma^{2}=\Gamma\circ \Gamma.$ 

\item Given two continuous semimartingales $X$ and $Y$, we define

\begin{eqnarray*}
L[X,Y]_{t}:=\mathbb{E}_{t}\left[\int^{T}_{t} \sigma_{u} d[X,Y]_{u}\right]
\end{eqnarray*}
and 

\begin{eqnarray*}
D[X,Y]_{t}:=\mathbb{E}_{t}\left[\int^{T}_{t} d[X,Y]_{u}\right],
\end{eqnarray*}
where the process $u\mapsto [X,Y]_{u}$, $u\in[0,T]$, is the quadratic covariation of the processes $X$ and $Y$.
\end{itemize}

\section{General expansion formulas}\label{sec:decomposition}

It is known that if the volatility process in a stochastic volatility model is independent of the asset price process 
(such models are called uncorrelated), then the following formula holds for a European style call option:

\begin{equation}\label{E:HW}
V_{t}=E_t [(BS)(t,X_{t},{\bar \sigma}_{t})].
\end{equation}
Here the symbol ${\bar \sigma}^2(t)$ stands for the average future variance defined by 
\begin{eqnarray*}
{\bar\sigma}^2_{t}:=\frac{1}{T-t} \int^{T}_{t}\sigma^{2}_{s}ds.
\end{eqnarray*}

The equality in (\ref{E:HW}) is called the Hull-White formula (see, e.g., \cite{F00}, page 51). For correlated models, that is, models where $\rho\neq 0$, there is a generalization of the Hull-White formula (see, e.g., formula (2.31) in \cite{F00}). However, the latter formula is significantly more complicated than the formula in (\ref{E:HW}).

Another way of generalizing the Hull-White formula was suggested in \cite{A06}. The idea used in \cite{A06} is to obtain an expansion of the random variable $V_{t}$ with the leading term equal to $E_t [(BS)(t,X_{t},{\bar \sigma}_{t})]$ and extra terms obtained using Malliavin calculus techniques. In \cite {A12}, a similar formula was found, in which the leading term contains the adapted projection of the future variance, that is, the quantity 
$$v^2_{t}:={\mathbb E}_t({\bar\sigma}^2_{t})=\frac{1}{T-t} \int^{T}_{t}{\mathbb E}_t[\sigma^{2}_{s}]ds,$$
instead of the future variance ${\bar\sigma}^2.$. The previous remark illustrates an important idea of switching from an anticipative process 
$t\mapsto{\bar\sigma}_{t}$ to a non-anticipative (adapted) process $t\mapsto v_{t}$. 
This idea was further elaborated in \cite{A12} in the case of the Heston model, which lead to a Hull-White type formula with the leading term equal to $(BS)(t,X_{t},v_t)$ and two more terms (see Theorem \ref{BS Deco15} below). In \cite{MV15}, the latter call price expansion was generalized to any stochastic volatility model. 

Let us define

$$M_{t}=\int^{T}_{0}\mathbb{E}_{t}\left[\sigma^{2}_{s}\right]ds.$$
It is not hard to see that the following equality holds:

$$dv^{2}_{t}= \frac{1}{T-t}\left[dM_{t}+\left(v^{2}_{t}-\sigma^{2}_{t}\right)dt\right].$$

The next assertion contains the above-mentioned decomposition formula due to Alòs. 

\begin{theorem}[BS expansion formula]\label{BS Deco15}
For every $t\in[0,T]$,
\begin{eqnarray}
V_{t}&=&(BS)(t,X_{t},v_{t}) \nonumber \\
&+&\frac{\rho}{2}\mathbb{E}_{t}\left[\int^{T}_{t}e^{-r(u-t)}\Lambda\Gamma(BS)(u,X_{u},v_{u})\sigma_{u}d[W,M]_{u}\right]
\nonumber \\
&+&\frac{1}{8}\mathbb{E}_{t}\left[\int^{T}_{t}e^{-r(u-t)}\Gamma^{2}(BS)(u,X_{u},v_{u})d[M,M]_{u}\right] \nonumber \\
&=&(BS)(t,X_{t},v_{t}) + (I) + (II).
\label{E:tri}
\end{eqnarray}
\end{theorem}

We will need the following statement established in \cite{MerinoPospisilSobotkaVives17}.

\begin{theorem}[General expansion formula]
\label{Generic Deco}
Let $B_{t}$, $t\in[0,T]$, be a continuous semimartingale with respect to the filtration $\mathcal{F}^W$, let $A(t,x,y)$ be a $C^{1,2,2} ([0,T]\times [0,\infty)\times [0,\infty))$ function, and let $v^2_t$ and $M_t$ be as above. 
Then, for every $t\in[0,T]$,

\begin{eqnarray*}
&&\mathbb{E}_{t}\left[e^{-r(T-t)}A(T,X_{T},v^{2}_{T})B_{T}\right]=A(t,X_{t},v^{2}_{t})B_{t}\\
&+&\mathbb{E}_{t}\left[\int^{T}_{t}e^{-r(u-t)}\partial_{y}A(u,X_{u},v^{2}_{u})B_{u}\frac{1}{T-u}\left(v^{2}_{u}-\sigma^{2}_{u}\right)du\right]\\
&+&\mathbb{E}_{t}\left[\int^{T}_{t}e^{-r(u-t)}A(u,X_{u},v^{2}_{u})dB_{u}\right]\\ 
&+&\frac{1}{2}\mathbb{E}_{t}\left[\int^{T}_{t}e^{-r(u-t)}\left(\partial^{2}_{x}-\partial_{x}\right)A(u,X_{u},v^{2}_{u})B_{u}\left(\sigma^{2}_{u}- v^{2}_{u}\right)du\right]\\ 
&+&\frac{1}{2}\mathbb{E}_{t}\left[\int^{T}_{t}e^{-r(u-t)}\partial^{2}_{y}A(u,X_{u},v^{2}_{u})B_{u}\frac{1}{(T-u)^{2}}d[M,M]_{u}\right]\\ 
&+&\rho \mathbb{E}_{t}\left[\int^{T}_{t}e^{-r(u-t)}\partial^{2}_{x,y}A(u,X_{u},v^{2}_{u})B_{u}\frac{\sigma_{u}}{T-u}d[W,M]_{u}\right]\\ 
&+&\rho \mathbb{E}_{t}\left[\int^{T}_{t}e^{-r(u-t)}\partial_{x}A(u,X_{u},v^{2}_{u})\sigma_{u} d[W,B]_{u}\right]\\ 
&+&\mathbb{E}_{t}\left[\int^{T}_{t}e^{-r(u-t)}\partial_{y}A(u,X_{u},v^{2}_{u})\frac{1}{T-u}d[M,B]_{u}\right].
\end{eqnarray*}
\end{theorem}

The following statement can be derived from Theorem \ref{Generic Deco}.

\begin{corollary}\label{BS Deco}
Suppose the functional $A_t:=A(t,X_{t},v^{2}_{t})$, $t\in[0,T]$, satisfies the Black-Scholes equation (\ref{FK}), and $B$ is adapted to the filtration $\mathcal{F}^{W}$. Then, for every $t\in[0,T]$,

\begin{eqnarray*}
e^{-r(T-t)}\mathbb{E}_{t}\left[A_{T}B_{T}\right]&=&A_{t} B_{t}\\
&+&\frac{\rho}{2}\mathbb{E}_{t}\left[\int^{T}_{t} e^{-r(u-t)} \Lambda\Gamma A_{u} B_{u} \sigma_{u} d[W,M]_{u}\right]\\
&+&\frac{1}{8}\mathbb{E}_{t}\left[\int^{T}_{t} e^{-r(u-t)} \Gamma^2 A_{u} B_{u} d[M,M]_{u}\right]\\
&+&\rho\mathbb{E}_{t}\left[\int^{T}_{t} e^{-r(u-t)} \Lambda A_{u}  \sigma_{u} d[W,B]_{u}\right]\\
&+&\frac{1}{2}\mathbb{E}_{t}\left[\int^{T}_{t} e^{-r(u-t)} \Gamma A_{u} d[M,B]_{u}\right]\\
&+&\mathbb{E}_{t}\left[\int^{T}_{t} e^{-r(u-t)} A_{u} dB_{u} \right].
\end{eqnarray*}
\end{corollary}

\begin{remark}\label{R:folo}
Theorem \ref{BS Deco15} follows from Corollary \ref{BS Deco} with $B\equiv 1$ and $A_t=(BS)(t,X_{t},v_{t}).$ 
\end{remark}

\begin{remark}\label{R:ro}
For the sake of shortness, the following notation will be used throughout the paper: 
$(BS)_t:=(BS)(t,X_{t},v_{t}),\quad t\in[0,T].$
\end{remark}

The terms $(I)$ and $(II)$ in formula (\ref{E:tri}) are not easy to evaluate. Therefore, it becomes important to find simpler approximations to 
$(I)$ and $(II)$ and estimate the error terms. We will next explain how to get an infinite expansion of the call price $V_{t}$, and in the next section, higher order approximations to $V_{t}$ will be obtained in the case of the Heston model.

The starting point in the construction of an infinite expansion of $V_{t}$ is the formula in (\ref{E:tri}). In \cite{A12}, Corollary \ref{BS Deco} was applied to the equality in (\ref{E:tri}). Only the main two terms in the expansion were kept \cite{A12}, while the remaining terms were ignored. The main idea used in the present paper is to apply Corollary \ref{BS Deco} to each new term, obtaining an infinite series with stochastic terms. By selecting which terms to keep in the approximation formula and which ones to discard, the approximation error can be controlled.

The process described above leads to the following expansion of $V_t$:
\begin{eqnarray}
V_{t}&=&  (BS)_{t} \label{BS High Order Approx} \nonumber \\
&+& \Lambda\Gamma (BS)_{t}\left(\frac{\rho}{2}L[W,M]_{t}\right)+\frac{1}{2} \Lambda^{2}\Gamma^{2} (BS)_{t}\left(\frac{\rho}{2}L[W,M]_{t}\right)^{2} \nonumber \\
&+& \Gamma^2 (BS)_{t}\left(\frac{1}{8}D[M,M]_{t}\right)+\frac{1}{2} \Gamma^4 (BS)_{t}\left(\frac{1}{8}D[M,M]_{t}\right)^{2} \nonumber \\
&+& \Lambda\Gamma^3 (BS)_{t}\left(\frac{\rho}{2}L[W,M]_{t}\right)\left(\frac{1}{8}D[M,M]_{t}\right) \nonumber \\
&+&  \ldots
\label{E:3}
\end{eqnarray}

\begin{remark}
In \cite{AGR}, an exact representation of $V_t$ is given in terms of a forest of iterated integrals, also called diamonds. The expansion of the call price found in the present section is equivalent to the one obtained in \cite{AGR}. 
\end{remark}

\section{Call price approximations in the Heston model}\label{sec:Heston}
The log-price process $X$ in the Heston model satisfies the following system of stochastic differential equations: 

\begin{eqnarray}
dX_{t}&=& \left(r-\frac{1}{2}\sigma^{2}_{t}\right)dt + \sigma_{t} \left(\rho dW_{t}  + \sqrt{1-\rho^{2}}d\widetilde{W}_{t} \right) \nonumber 
\\
d\sigma^{2}_{t}&=& \kappa \left(\theta - \sigma^{2}_{t}\right)dt + \nu \sqrt{\sigma^{2}_{t}}dW_{t}. \nonumber
\end{eqnarray}
Here the process $\sigma^2$ models the stochastic variance of the asset price, $\theta> 0$ is the long-run mean level of the variance, $\kappa> 0$ is the rate at which $\sigma_{t}$ reverts to the mean $\theta$, $\nu> 0$ is the volatility of volatility parameter, and $r\ge 0$ is the interest rate. The initial conditions for the volatility process $\sigma$ and the log-price process $X$ will be denoted by $\sigma_0> 0$ and $x_0$, respectively. We will assume throughout the rest of the paper that the Feller condition $2\kappa\theta \geq \nu^{2}$ (the non-hitting condition) is satisfied. 

In this section, the general results established in Section \ref{sec:decomposition} are used to obtain new approximation formulas for the call price in the Heston model. Our results generalize and sharpen the approximation formula obtained in \cite{A12} and \cite{ADV}, providing more terms in the small vol-vol asymptotic expansion of the call price. Moreover, the error terms in our formulas are of higher order than the error term of the form $O(\nu^{2}\left(\left|\rho\right|+\nu\right)^{2})$ appearing in \cite{A12} and \cite{ADV}. The proofs of these results will be given in the next subsection.

We start with an assertion that provides an approximation of the order $O(\nu^{3}(\left|\rho\right|+\nu)).$

\begin{theorem}[2nd order approximation formula]\label{BS 2nd Approximation}
For every $t\in[0,T]$,
\begin{eqnarray*}
V_{t}&=&(BS)(t,X_{t},v_{t})\\
&+&\Gamma^{2} (BS)(t,X_{t},v_{t})\left(\frac{1}{8}D[M,M]_{t}\right)\\
&+&\Lambda \Gamma (BS)(t,X_{t},v_{t})\left(\frac{\rho}{2}L[W,M]_{t}\right)\\
&+&\frac{1}{2}\Lambda^{2} \Gamma^{2} (BS)(t,X_{t},v_{t})\left(\frac{\rho}{2}L[W,M]_{t}\right)^{2}\\ 
&+&\rho\Lambda^{2} \Gamma (BS)(t,X_{t},v_{t})L[W,\frac{\rho}{2}L[W,M]]_{t}\\
&+&\epsilon_{t},
\end{eqnarray*}
where $\epsilon_{t}$ is the error term satisfying 
\begin{eqnarray*}
|\epsilon_{t}| \leq \nu^{3}\left(\left|\rho\right| +\left|\rho\right|^3 + \nu\right)\left(\frac{1}{r}\wedge (T-t)\right)\Pi(\kappa, \theta),
\end{eqnarray*}
and $\Pi(\kappa, \theta)$ is a positive constant depending on $\kappa$ and $\theta$.
\end{theorem}

The next assertion contains an approximation formula with the error term of the form $O(\nu^{4} (1+|\rho|).$

\begin{theorem}[3rd order approximation formula]\label{BS Approximation2}
For every $t\in[0,T]$,
\begin{eqnarray*}
V_{t}&=&  (BS)_{t} \\
&+& \Lambda\Gamma (BS)_{t}\left(\frac{\rho}{2}L[W,M]_{t}\right) +\frac{1}{2} \Lambda^{2}\Gamma^{2} (BS)_{t}\left(\frac{\rho}{2}L[W,M]_{t}\right)^{2}\\
&+& \frac{1}{6}\Lambda^{3}\Gamma^3 (BS)_{t}\left(\frac{\rho}{2}L[W,M]_{t}\right)^{3}+\Lambda\Gamma^3 (BS)_{t}\left(\frac{\rho}{2}L[W,M]_{t}\right)\left(\frac{1}{8}D[M,M]_{t}\right)\\
&+& \rho \Lambda^{2}\Gamma (BS)_{t} L[W,\frac{\rho}{2}L[W,M]]_{t} +\rho \Lambda \Gamma^2 (BS)_{t} L[W,\frac{1}{8}D[M,M]]_{t}\\
&+&\frac{1}{2} \Lambda\Gamma^{2} (BS)_{t}D[M,\frac{\rho}{2} L[W,M]]_{u}\\
&+& \rho \Lambda^{3}\Gamma^2 (BS)_{t}\frac{\rho}{2}L[W,M]_{t}L[W,\frac{\rho}{2}L[W,M]]_{t}\\
&+&\rho \Lambda^{3}\Gamma (BS)_{t}L[W,\rho L[W,\frac{\rho}{2}L[W,M]]]_{t}\\
&+&\Gamma^2 (BS)_{t}\left(\frac{1}{8}D[M,M]_{t}\right)\\
&+&\epsilon_{t}.
\end{eqnarray*}
where $\epsilon_{t}$ is the error term satisfying
\begin{eqnarray*}
|\epsilon_{t}| \leq \nu^{4}\left(1+ \rho^{2}\left(1+\rho^{2}\right) + \left|\rho\right|\nu\left(1+\rho^{2}\right)\right)\left(\frac{1}{r}\wedge (T-t)\right)\Pi(\kappa, \theta),
\end{eqnarray*}
and $\Pi(\kappa, \theta)$ is a positive constant depending on $\kappa$ and $\theta$.
\end{theorem}

For the uncorrelated Heston model, we obtain a similar expansion with fewer terms and a better error estimate. 

\begin{corollary}\label{BS Approximation zero-corr}
If $\rho=0$, then
\begin{eqnarray*}
V_{t}&=&  (BS)_{t} \\
&+& \Gamma^2 (BS)_{t}\left(\frac{1}{8}D[M,M]_{t}\right) + \frac{1}{2}\Gamma^4 (BS)_{t}\left(\frac{1}{8}D[M,M]_{t}\right)^{2}\\
&+& \frac{1}{2}\Gamma^3 (BS)_{t}D[M,\frac{1}{8}D[M,M]]_{t}\\
&+& \epsilon_{t}
\end{eqnarray*}
where $\epsilon_{t}$ is the error term satisfying
\begin{eqnarray*}
|\epsilon_{t}| \leq \nu^{6}\left(\frac{1}{r}\wedge (T-t)\right)\Pi(\kappa, \theta),
\end{eqnarray*}
and $\Pi(\kappa, \theta)$ is positive constant depending on $\kappa$ and $\theta$.
\end{corollary}
\begin{remark}
Call price approximations similar to those formulated above in the case of the Heston model can also be obtained for the Bates model 
(see \cite{MerinoPospisilSobotkaVives17} for more information on call price approximations in the Bates model).
\end{remark}

\section{Numerical results}\label{SS:numa}
In this section, we compare the performance of the call price approximation formula proposed in \cite{A12} and \cite{ADV} with that of the new approximation formulas obtained in this paper. To simplify the notation we call the formula obtained in \cite{A12} and \cite{ADV} the formula with error order $O(\nu^2)$, while the two formulas obtained in the present paper will be called the formulas with error orders $O(\nu^3)$ and $O(\nu^4)$, respectively (see Theorems \ref{BS 2nd Approximation} and \ref{BS Approximation2}). We also make a similar comparison in the zero-correlation case. Here we compare the formula with error order $O(\nu^4)$ established in \cite{A12} and \cite{ADV} and the new formula that has an error of order $O(\nu^6)$ found in the present paper (see Corollary \ref{BS Approximation zero-corr}). As a benchmark price, we choose a call price obtained using a Fourier transform based pricing formula. This is one of the standard approaches to pricing European options under stochastic volatility models. In particular, we use a semi-closed form solution with one numerical integration as a reference price (see \cite{MrazekPospisil17})\footnote{With a slight modification mentioned in \cite{Ga} in order not to suffer from the "Heston trap" issues.}. The comparison between approximations is made with two important aspects in mind: the practical precision of the pricing formula and the efficiency of the formula expressed in terms of the computational time needed for particular pricing tasks. 

Our next goal is to illustrate the quality of approximation of the call price in the Heston model for various values of $\rho$ and $\nu$, while keeping the other parameters fixed\footnote{We choose the following values of the parameters: $S_0=100$; $r=0.001$; $v_0=0.25$; $\kappa=1.5$; $\theta=0.2$.}. We understand the error in the price as the relative error in a $\log_{10}$ scale.

\begin{figure}
\includegraphics[height=8cm,width=0.95\textwidth]{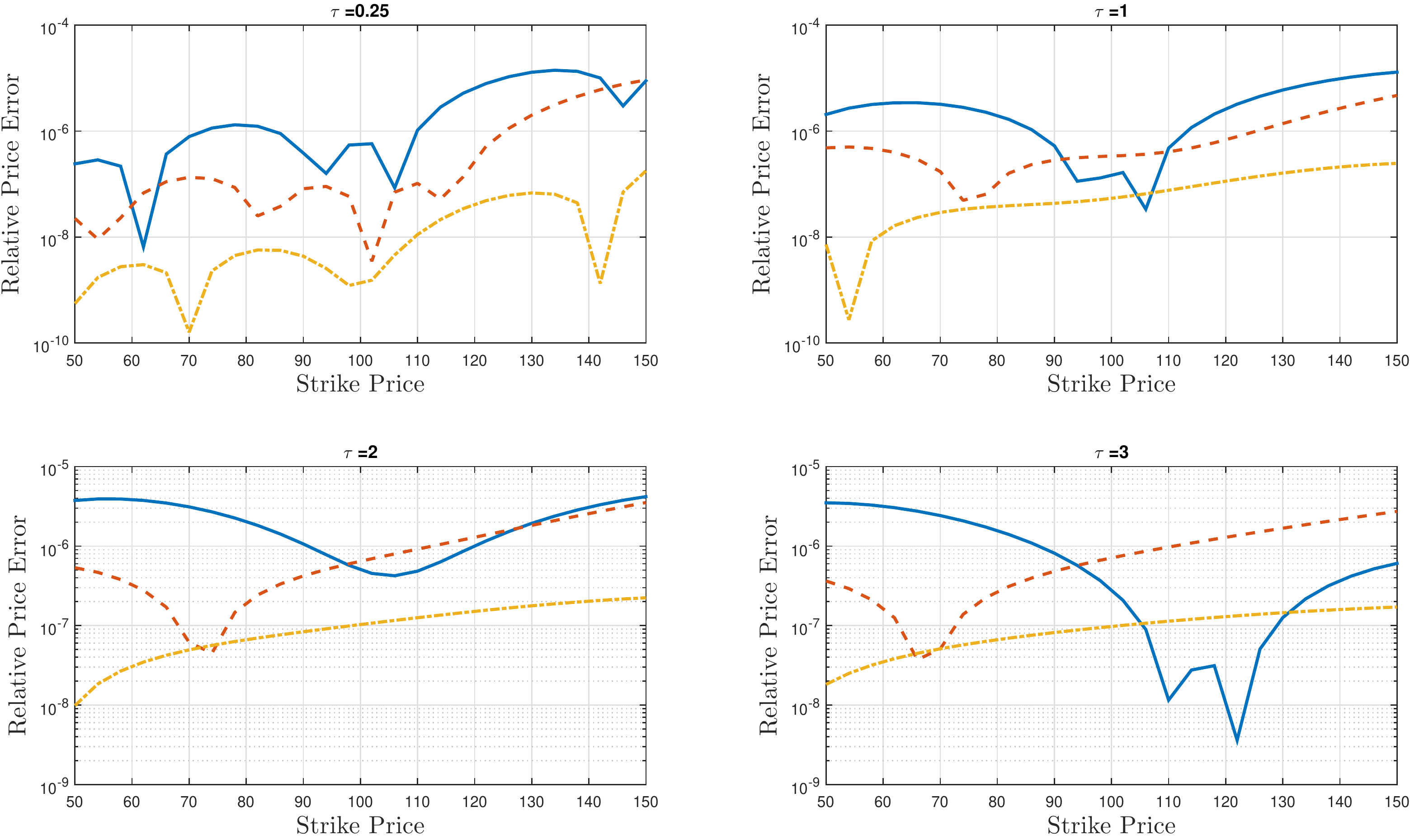}
\caption{Comparison of the three different approximation formulas and the reference prices. The figure represents the relative price error in a 
$\log_{10}$ scale in terms of the option strikes at four different slices of time. The blue line illustrates the approximation with the error 
term of order $O(\nu^{2})$, the red line is the approximation the error term of order $O(\nu^{3})$, while the yellow line corresponds to the approximation with the error term of order $O(\nu^4)$. The parameters are $\rho=-0.2$, $\nu= 5\% $, $S_0=100$, $r=0.001$,  $v_0=0.25$, $\kappa=1.5$, and $\theta=0.2$.  }
\label{Fig:lowNuRho}
\end{figure}

Figure \ref{Fig:lowNuRho} shows the call price approximations in the case, where the vol-vol parameter $\nu$ and the absolute value of the instantaneous correlation $\rho$ are both small. We observe that in this case, the approximation formula with the error of order $O(\nu^3)$ in general performs better than the formula, in which the error order is $O(\nu^2)$. However, there are exceptions in some OTM cases for $\tau=3$. The call price approximation with the error of order $O(\nu^4)$ is much better when the error is around 
$10^{-7}-10^{-10}$.

\begin{figure}
\includegraphics[height=8cm,width=0.95\textwidth]{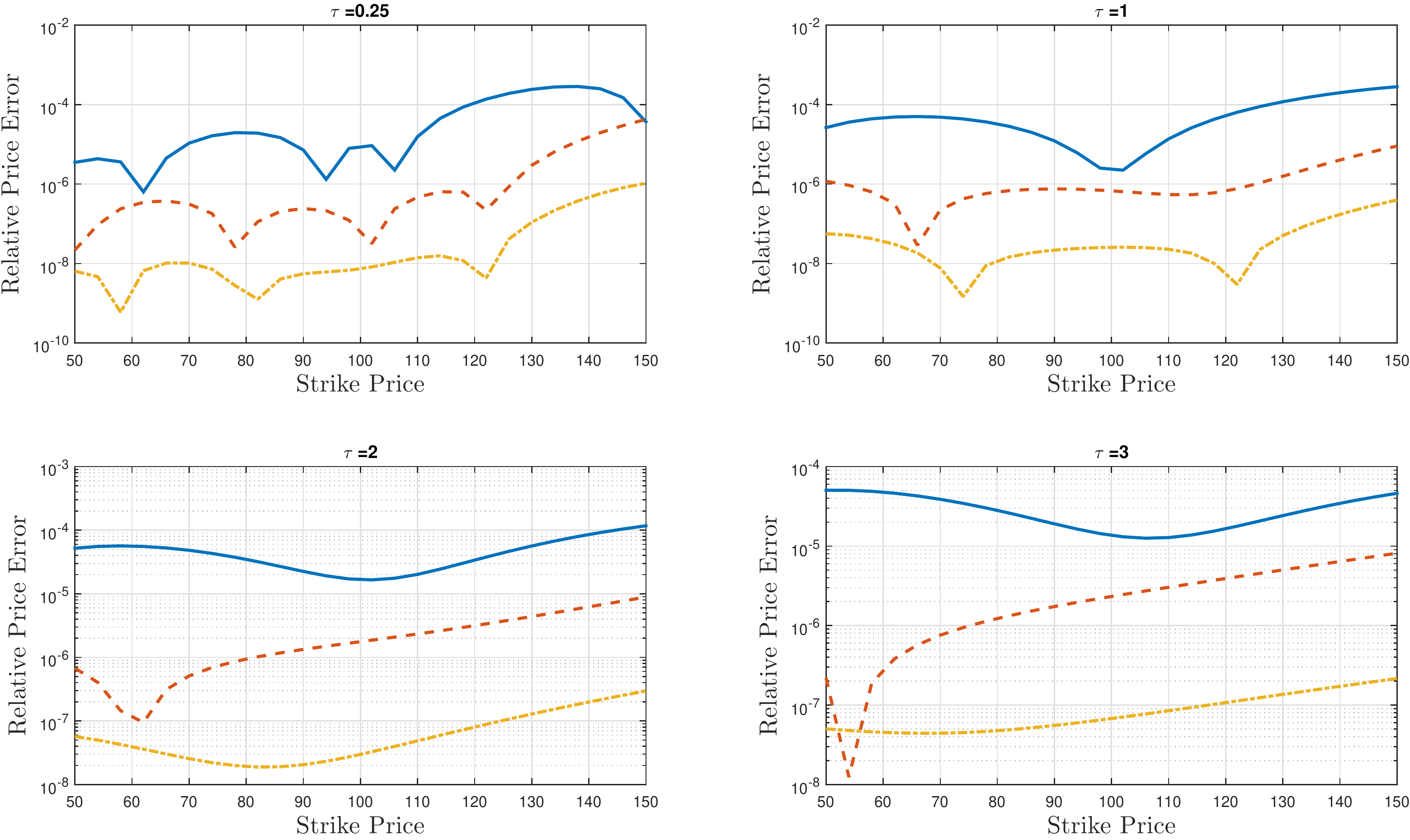}
\caption{Comparison of the three different approximation formulas and the reference prices. The figure represents the relative price error in a 
$\log_{10}$ scale in terms of the option strikes at four different slices of time. The blue line is the approximation with the error 
term of order $O(\nu^{2})$, the red line represents the approximation with the error term of order $O(\nu^{3})$, while the yellow line is the approximation with the error term of order $O(\nu^4)$. The parameters are $\rho=-0.8$, $\nu= 5\% $, $S_0=100$, $r=0.001$,  $v_0=0.25$, $\kappa=1.5$, and $\theta=0.2$.  }
\label{Fig:LowNuHighRho}
\end{figure}

In Figure \ref{Fig:LowNuHighRho}, we deal with the case, where $\nu$ is small while $|\rho|$ is close to one. We observe that the new approximation formulas perform clearly better than the previously known one. The approximation error is in the range $10^{-4}-10^{-8}$ for the formula, in which the error term is of order 
$O(\nu^3)$, and for the formula with the error term of order $O(\nu^4)$, the approximation error is in the range  
$10^{-7}-10^{-10}$. 

\begin{figure}
\includegraphics[height=8cm,width=0.95\textwidth]{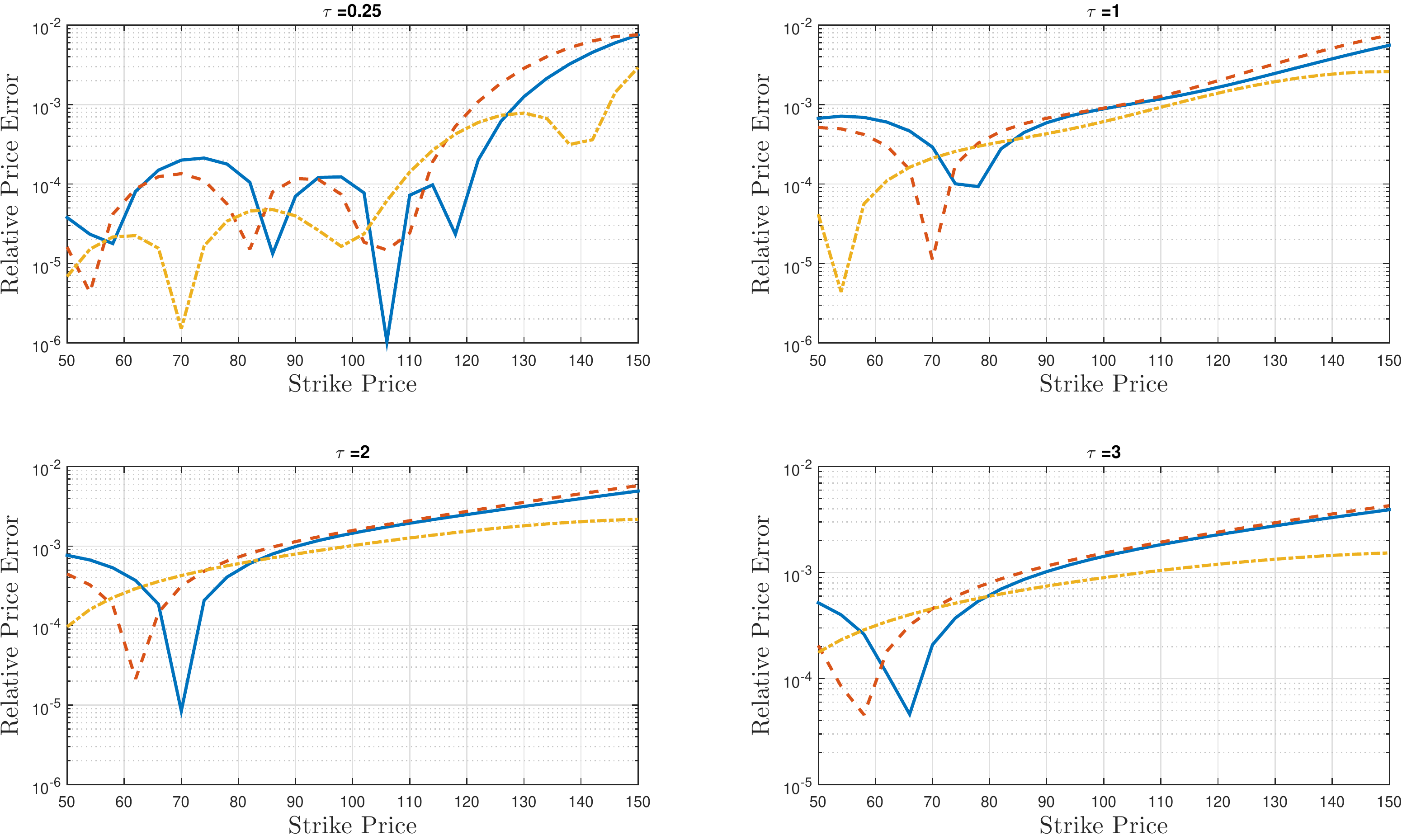}
\caption{Comparison of the three different approximation formulas and the reference prices. The figure represents the relative price error in a 
$\log_{10}$ scale in terms of the option strikes at four different slices of time. The blue line is the approximation with the error 
term of order $O(\nu^{2})$, the red line is the approximation with the error term of order  $O(\nu^{3})$, and the yellow line is the approximation with the error term of order $O(\nu^4)$. The parameters are $\rho=-0.2$, $\nu= 50\% $, $S_0=100$, $r=0.001$,  $v_0=0.25$, $\kappa=1.5$, and $\theta=0.2$.  }
\label{Fig:HighNuLowRho}
\end{figure}

Figure \ref{Fig:HighNuLowRho} concerns the case of high volatility $\nu$ and small absolute value of $\rho.$ Her we observe that the three approximation formulas show a similar performance. The approximation formula, where the error order is $O(\nu^4)$ seems to perform a little better, but not by much.

\begin{figure}
\includegraphics[height=8cm,width=0.95\textwidth]{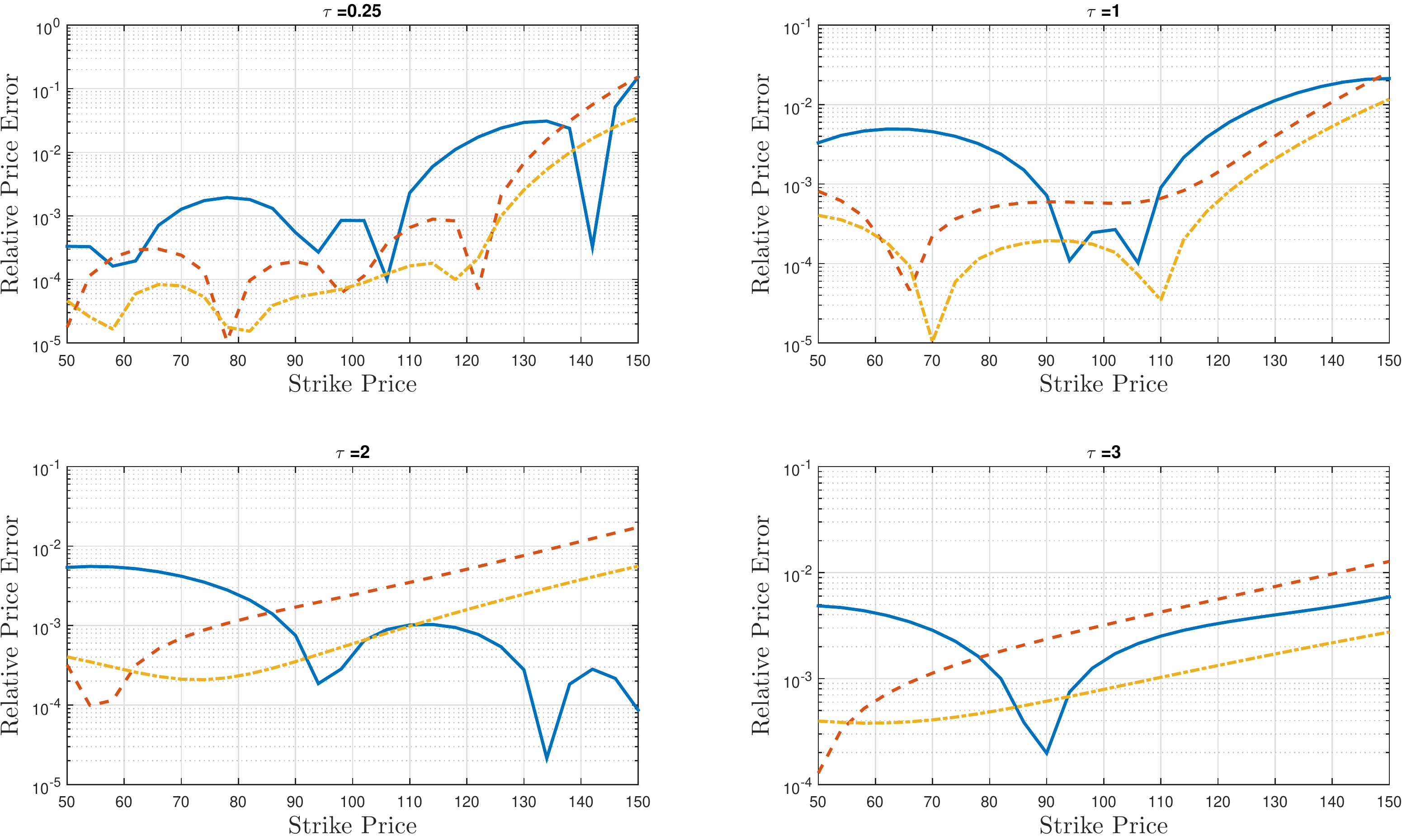}
\caption{Comparison of the three different approximation formulas and the reference prices. The figure represents the relative price error in a 
$\log_{10}$ scale in terms of the option strikes at four different slices of time. The blue line is the approximation with the error 
term of order $O(\nu^{2})$, the red line represents the approximation with the error term of order $O(\nu^{3})$, while the yellow line is the approximation with the error term of order  $O(\nu^4)$. The parameters are $\rho=-0.8$, $\nu= 50\% $, $S_0=100$, $r=0.001$,  $v_0=0.25$, $\kappa=1.5$, and $\theta=0.2$.  }
\label{Fig:highNURho}
\end{figure}

Figure \ref{Fig:highNURho} illustrates the performance of the formulas when both parameters are not well suited for the approximation, e.g.,
when $\nu = 50\%$ and $\rho=-0.8$. The resulting errors are depicted in Figure \ref{Fig:highNURho}. Here we observe that the approximations have a similar quality. The approximation formula with the error term of order $O(\nu^4)$ seems to perform better than the other formulas, while the formula with the error term of order $O(\nu^3)$ performs better only in the short time interval.

We have already observed that the approximation formulas obtained in the present paper perform better than the previously known formula when 
$|\rho|$ is close to one, while $\nu$ is small. On the other hand, the improvement in the performance is not significant if $\nu$ is large. In order to show that the sharpness of the approximation can also be improved in the case of large values of the parameter $\nu$, we compare the benchmark price with its approximations in the zero-correlation case.

\begin{figure}
\includegraphics[height=8cm,width=0.95\textwidth]{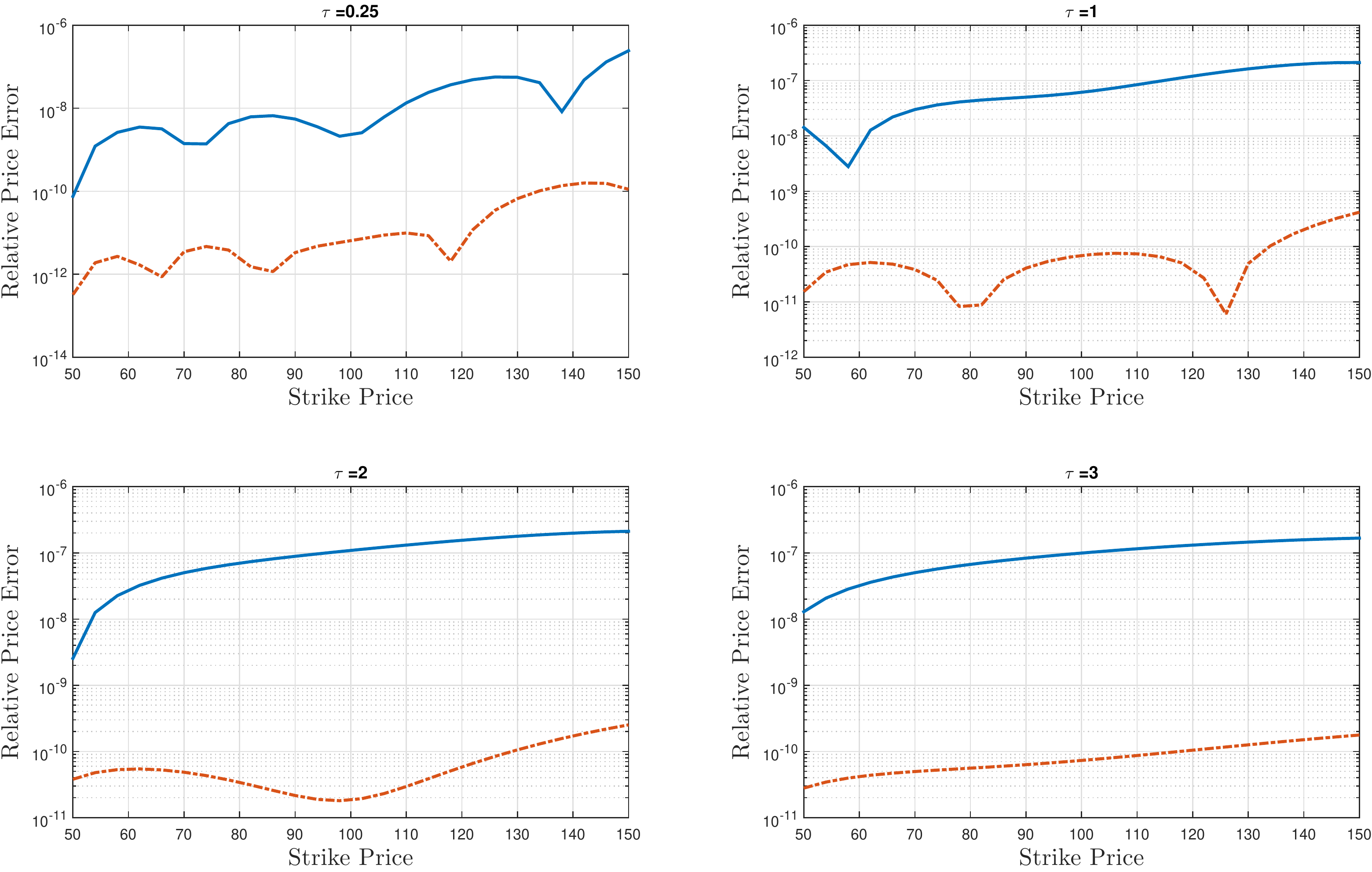}
\caption{Comparison of the two different approximation formulas and the reference prices. The figure represents the relative price error in a 
$\log_{10}$ scale in terms of the option strikes at four different slices of time. The blue line is the approximation with the error 
term of order $O(\nu^{4})$, while the red line is the approximation with the error term of order $O(\nu^{6})$. The parameters are $\rho=0$, $\nu= 5\% $, $S_0=100$; $r=0.001$;  $v_0=0.25$; $\kappa=1.5$; $\theta=0.2$.  }
\label{Fig:ZCLowNu}
\end{figure}

\begin{figure}
\includegraphics[height=8cm,width=0.95\textwidth]{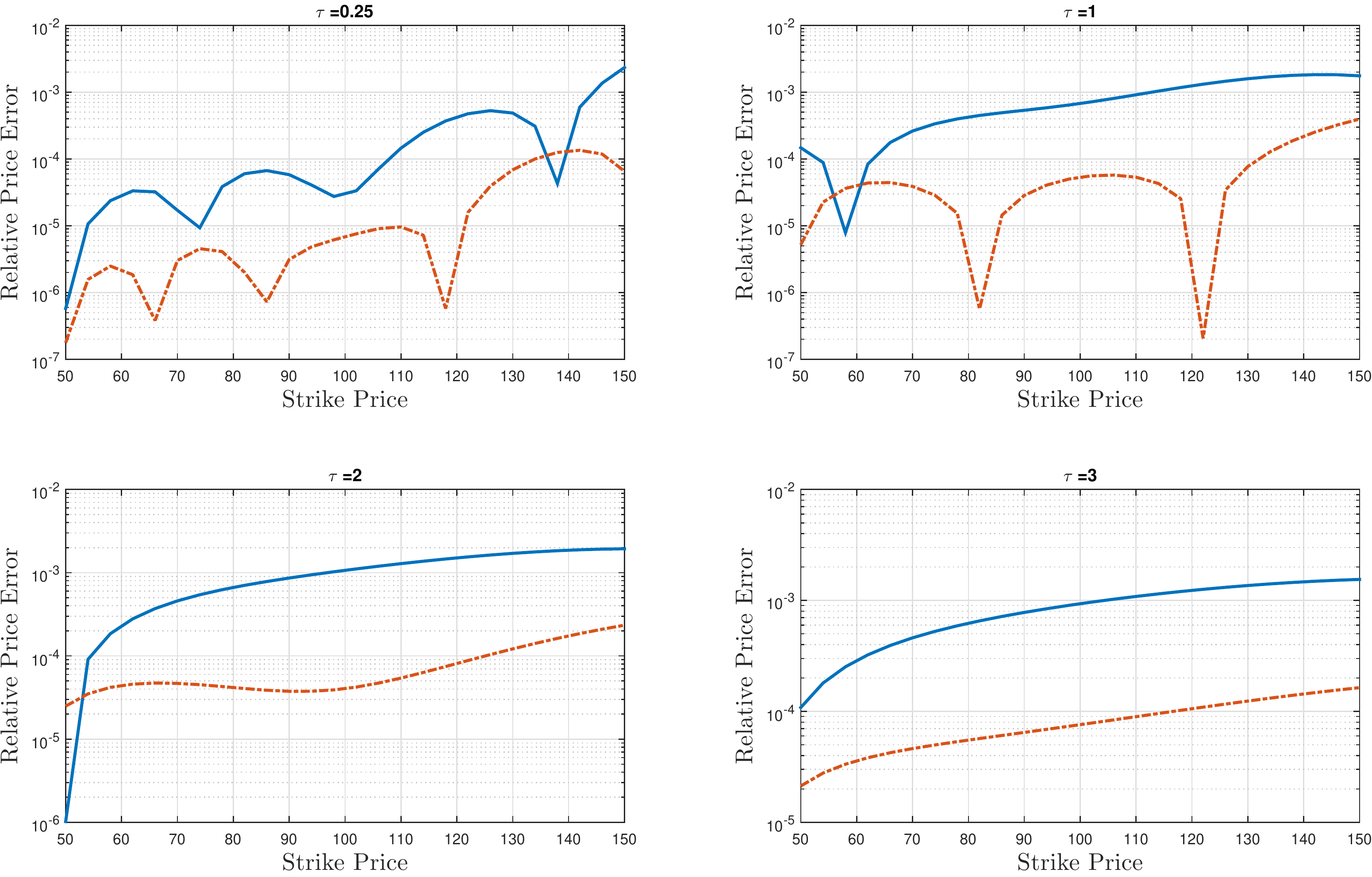}
\caption{Comparison of the two different approximation formulas and reference prices. The figure represents the relative price error in a 
$\log_{10}$ scale in terms of the option strikes at four different slices of time. The blue line is the approximation with the error 
term of order $O(\nu^{4})$, while the red line is the approximation with the error 
term of order $O(\nu^{6})$. The parameters are $\rho=0$, $\nu= 50\% $, $S_0=100$, $r=0.001$,  $v_0=0.25$, $\kappa=1.5$, and $\theta=0.2$.  }
\label{Fig:ZCHighNu}
\end{figure}

In Figure \ref{Fig:ZCLowNu}, we illustrate the case of low volatility of volatility. Both $O(\nu^2)$ and $O(\nu^3)$ formulas perform rather well with a very small error, but the new approximation behaves much better than the known one. Figure \ref{Fig:ZCHighNu} shows the approximations in the uncorrelated Heston model when $\nu$ is large. We can see that the approximation formula of order $O(\nu^{6})$ obtained in this paper performs much better than the previously known formula, especially in the long-term scenario.
\newpage

\newpage

One of the main advantages of the proposed pricing approximations is their computational efficiency. To compare the amount of time spent on computations, three pricing tasks were set up. We used a batch of $100$ various call options with different strikes and times to maturity, including OTM, ATM, and ITM options with short-, mid- and long-term times to maturity. The first task was to evaluate prices of the options in the batch with respect to $100$ (uniformly) randomly sampled parameter sets. This needed a similar number of price evaluations as in a market calibration task with a very good initial guess. Further on, we repeated the same trials  for $1.000$ and $10.000$ parameter sets to mimic the number of evaluations for a typical local-search calibration and a global-search calibration, respectively (for more information about calibration tasks see e.g. \cite{MikhailovNogel03} and \cite{MrazekPospisilSobotka16ejor}).  

Our results are listed in Table \ref{Tab:Res1}. The call prices were analytically calculated in all the cases. We observe 
that for the trials of $100$ and $1.000$ sets, the amounts of time spent on computations are rather similar. For the trial of $10.000$ sets, the experiment based on the
approximation with the error term of order $O(\nu^4)$ was a little bit slower than the other experiments. 

\begin{table}[ht]
\caption{Efficiency of the Heston call price approximations}\label{Tab:Res1}
\centering
\begin{tabular}{lcrc}
Pricing approach    & Task & Time$^\dagger$ [sec]  & Speed-up factor\\
\toprule
\multirow{3}{*}{Heston-Lewis}     &  $\#1$    & $3.63$ &   -  \\
          & $\# 2$         & $33.52$  &   - \\
          & $\# 3$     & $336.59$  &   -\\
\midrule
\multirow{3}{*}{Approximation of order $O(\nu^{2})$}     &  $\#1$    & $0.08$ &   $45\times$  \\
          & $\# 2$         & $0.76$  &   $44\times$  \\
          & $\# 3$     & $7.41$  &   $45\times$ \\
\midrule
\multirow{3}{*}{Approximation of order $O(\nu^{3})$}     & $\# 1$    & $0.08$  & $45\times$  \\
          & $\# 2$         & $0.78$ &     $43\times$ \\
          & $\# 3$     & $7.77$  &   $43\times$ \\
\midrule
\multirow{3}{*}{Approximation of order $O(\nu^{4})$}     & $\# 1$    & $0.10$  & $36\times$  \\
          & $\# 2$         & $0.91$ &     $37\times$\\
          & $\# 3$     & $8.87$  &   $38\times$ \\					
\bottomrule
\end{tabular}\\
$^\dagger$ {\footnotesize The results were obtained on a PC with Intel Core i7-7700HQ CPU @2.80 GHz 2.80GHz and 16 GB RAM.}
\end{table}

The table shows that the approximations, where the error order is $O(\nu^{2})$ or $O(\nu^3)$ are around 43-45 times faster than the approximation based on the fast Fourier transform methodology, while the approximation with the error of order $O(\nu^4)$ is around 36 times faster than the latter one. Therefore, the approximations with error order $O(\nu^{2})$ or $O(\nu^3)$ are around 1.14-1.25 times less time-consuming than the $O(\nu^{4})$-approximation. 

\section{Conclusion.}\label{sec:conclusion}
In this paper, we find various expansions for call prices in general stochastic volatility models. In the special case of the Heston model, we derive sharp approximation formulas with error estimates for the call price. These formulas have a higher order of accuracy than the previously known ones. In Section \ref{sec:Heston}, an exact second order approximation formula is established for the call price in the Heston model that has an error term of order $O(\nu^{3}(\left|\rho\right| +\nu))$, where $\nu$ is the vol-vol parameter and $\rho$ is the correlation coefficient. We also find a sharper formula with the error term of order $O(\nu^{4}(1+\rho^2))$ and seven additional significant terms. In Section \ref{SS:numa}, the numerical performance of the approximation formulas obtained in the present paper is illustrated. We observe that the new formulas are very efficient for low values of the vol-vol parameter $\nu$, or when the time to maturity is small. We also observe that for the uncorrelated Heston model, the number of terms, which have to be taken into account in computations, decreases substantially. For the call price in the uncorrelated Heston model, we find an approximation that has an error term of order $O(\nu^{6})$. The approximations to the call price obtained in the present paper are computationally more efficient than those proposed in \cite{A12} and \cite{ADV}. 

\begin{appendices}

\section{Auxiliary lemmas}

We will need several results from \cite{ADV} in the proof. The following notation will be used in the sequel:
$$\varphi(t):=\int^{T}_{t} e^{-\kappa(z-t)} dz = \frac{1}{\kappa}\left(1-e^{-\kappa(T-t)}\right).$$
We will also use the following processes: 
\begin{align}
&U_{t}=\frac{\rho}{2}L[W,M]_{t}\quad\mbox{and}\quad R_{t}=\frac{1}{8}D[M,M]_{t}
\end{align}
(see \cite{A12}, \cite{ADV}, and \cite{MerinoPospisilSobotkaVives17}). 


The following lemma (see \cite{A12}) can be used to get error estimates in call price approximation formulas. 

\begin{lemma}\label{fitage} 
Let $0\leq t\leq s\leq T$ and $\mathcal{G}_{t}:=\mathcal{F}_{t}\vee
\mathcal{F}_{T}^{W}.$ Then, for every integer $n\geq 0,$ there exists $C=C(n)> 0 $ such that
\begin{equation*}
\left\vert \mathbb{E} \left( \left. {\Lambda^{n}\Gamma (BS)}\left( s,X_{s},v_{s}\right)
\right\vert \mathcal{G}_{t}\right) \right\vert \leq C \left( \int_{s}^{T}\mathbb{E}_s
\left(\sigma_{\theta }^{2}\right) d\theta \right)^{-\frac{1}{2}\left(
n+1\right) }.
\end{equation*}
\end{lemma}

The next statement holds true for the Heston model (see \cite{ADV}). 
\begin{lemma}\label{remarkADV15}
The following are valid: 

\begin{enumerate}
\item If $s\geq t$, then
\begin{equation*}
E_t(\sigma^2_s)=\theta+(\sigma_t^2-\theta)e^{-\kappa (s-t)}=\sigma_t^2 e^{-\kappa (s-t)}+\theta (1-e^{-\kappa (s-t)}).
\end{equation*}
In particular, the previous quantity is bounded from below by $\sigma_t^2\wedge\theta$ and from above by $\sigma_t^2\vee\theta$.

\item 
$E_t\left(\int_{t}^{T}\sigma _{s}^{2}ds\right) = \theta\left(T-t\right) + \frac{\sigma_{t}^{2}-\theta}{\kappa}\left(1-e^{-\kappa\left( T-t\right) }\right).$

\item 
$dM_t=\nu\sigma_t\varphi(t)dW_t.$

\item 
$\frac{\rho}{2}L[W,M]_{t}:=\frac{\rho}{2} E_t\left(\int_{t}^{T}\sigma _{s}d \left[M,W\right]_{s}\right)=\frac{\rho}{2}\nu\int_{t}^{T}E_t \left(\sigma
_{s}^{2}\right) \varphi(s)ds.$

\item 
$\frac{1}{8}D[M,M]_{t}:=\frac{1}{8}E_t\left(\int_{t}^{T}d \left[M,M\right]_s\right) \
=\ \frac{1}{8}\nu ^{2}\int_{t}^{T}E_t\left(\sigma _s^2\right)
\varphi(s)^2ds.$

\item 
$\frac{\rho}{2}dL[W,M]_{t} = \frac{\rho \nu^2}{2}\left(\int_t^T e^{-\kappa(z-t)}\varphi(z)dz\right)\sigma_t dW_t-\frac{\rho\nu}{2}\varphi(t)\sigma^2_t dt$,

\item 
$\frac{1}{8}dD[M,M]_{t} = \frac{\nu^3}{8}\left(\int_t^T e^{-\kappa(z-t)}\varphi(z)^2dz\right) \sigma_t dW_t-\frac{\nu^2}{8}\varphi(t)^2 \sigma_t^2 dt$.
\end{enumerate}
\end{lemma}

In the following statement, lower bounds are provided for the adapted projection of the future variance.

\begin{lemma}\label{tercerlema} For every $s\in[0,T]$,\\ \\
\begin{tabular}{cl}
(i)  &   $\int_s^T E_s(\sigma_u^2) du\  \geq\  \frac{\theta\kappa}{2}\varphi(s)^2$, \\ \\
(ii) &  $\int_{s}^{T}E_{s}\left( \sigma _u^2\right) du\  \geq\  \sigma _{s}^{2}\varphi(s)$.
\end{tabular}
\end{lemma}

The next lemma is used in the proof of Theorem \ref{BS 2nd Approximation} and numerical computations related to it.

\begin{lemma}\label{Lemma_LUW}
The following formulas hold: 
\begin{eqnarray*} 
L[W,L[W,M]]_{t}&=&\mathbb{E}_{t}\left[\int^{T}_{t} \sigma_{u} d[L[W,M],W]_{u}\right]\\
&=& \nu^2 \int^{T}_{t} \mathbb{E}_{t}\left[\sigma^{2}_{u}\right] \left(\int_u^T e^{-\kappa(z-u)}\varphi(z)dz\right) du\\
&=& \frac{\nu^2}{2\kappa^3}\left\{2\left[\sigma^2 + \theta\left(\kappa\tau -3\right)\right]\right. \\
 &+& \left. e^{-\kappa\tau}\left[\theta\left(\kappa^{2}\tau^{2} + 4\kappa\tau + 6\right) - \sigma^{2}\left(\kappa^{2}\tau^{2} + 2 \kappa \tau +2\right)\right]\right\}
\end{eqnarray*}
and
\begin{eqnarray*}
dL[W,L[W,M]]_{t}&=& \nu^3\left[\int^{T}_{t}\left(\int^{T}_{u} e^{-\kappa(z-u)}\varphi(z)dz\right)e^{-\kappa(u-t)} du\right]\sigma_{t}dW_{t}\\
&-& \nu^2\left(\int^{T}_{t} e^{-\kappa(z-t)}\varphi(z)dz\right)\sigma^{2}_{t} dt.\\
\end{eqnarray*}
\end{lemma}

Similarly, the next lemma is used in the proof of Theorem \ref{BS Approximation2} and numerical computations related to it.

\begin{lemma}\label{L:1}
The following equalities hold true in the Heston model:
\begin{eqnarray*}
D[M,\frac{1}{8}D[M,M]]_{t}&=&\frac{\nu^3}{8} \int^{T}_{t} \mathbb{E}_{t}\left[\sigma^{2}_{u}\right] \left(\int_t^T e^{-\kappa(z-u)}\varphi(z)^2dz\right) du\\
&=&\frac{\nu^3 }{16 \kappa^4}\left\{\theta\left[\left(2\kappa\tau -7\right)+2e^{-\kappa\tau}\left(\kappa^{2}\tau^{2}+2\kappa\tau +4\right)-e^{-2\kappa\tau}\right]\right. \\
 &-& \left.2\sigma^2 e^{-\kappa\tau}\left[\kappa^2 \tau^2 - 2 cosh\left(\kappa\tau\right)+2\right]\right\},
\end{eqnarray*}
\begin{eqnarray*}
D[M,L[W,M]]_{t}&=&\nu^3\int^{T}_{t}\mathbb{E}_{t}\left[\sigma^{2}_u\right] \varphi(u)\left(\int_t^T e^{-\kappa(z-t)}\varphi(z)dz\right)du\\
&=&\frac{\nu^3}{4\kappa^4}\left\{\theta\left[2\kappa\tau e^{-2\kappa\tau}+\left(4\kappa\tau-13\right)+2e^{-\kappa\tau}\left(\kappa^{2}\tau^{2}+6\kappa\tau+4\right)+5e^{-2\kappa\tau}\right]\right. \\
 &+& \left.2\sigma^2\left[e^{-\kappa\tau}\left(-\kappa^{2}\tau^{2}-4\kappa\tau+2\right)-2e^{-2\kappa\tau}\left(\kappa\tau+2\right)+2\right]\right\},\\
\end{eqnarray*}
and
\begin{eqnarray*}
L[W,L[W,L[W,M]]]_{t}&=&\nu^3\int^T_t \mathbb{E}_{t}\left[\sigma^{2}_{u}\right] \left(\int^{T}_{u}\left(\int^{T}_{s} e^{-\kappa(z-s)}\varphi(z)dz\right)e^{-\kappa(s-u)} ds\right) du\\
&=&\frac{\nu^3  e^{-\kappa \tau} }{6 \kappa^4} \left(\theta  \left(6 e^{\kappa  \tau} (\kappa  \tau-4)+\kappa^3\tau^3 + 6\kappa^2\tau^2+ 18 \kappa\tau+24\right)\right.\\
&+& \left.\sigma ^2 \left(6e^{\kappa \tau}-6-\kappa^3\tau^3 - 3\kappa^2\tau^2 - 6 \kappa \tau\right)\right).
\end{eqnarray*}
\end{lemma}

The statement formulated below is used in the proof of Corollary \ref{BS Approximation zero-corr}. It is assumed on it that the correlation coefficient in the Heston model is equal to zero. 

\begin{lemma}
The following equality holds in the Heston model:
\begin{eqnarray*}
D[M,\frac{1}{8}D[M,M]]_{t}&=&\frac{\nu^4}{8}\int^{T}_{t}\mathbb{E}_{t}\left[\sigma^{2}_u\right] \varphi(u)\left(\int_u^T e^{-\kappa(z-u)}
\varphi(z)^2dz\right) du\\
&=& \frac{\nu^4 }{48\kappa^5}\left\{\theta\left[3e^{-\kappa\tau}\left(2\kappa^2 \tau^2+6\kappa\tau+5\right)+6e^{-2\kappa\tau}
\left(\kappa\tau+1\right)+\left(6\kappa\tau-22\right)+e^{-3\kappa\tau}\right]\right. \\
&+& \left.3\sigma^{2}\left[e^{-\kappa\tau}\left(-2\kappa^2 \tau^2-2\kappa\tau+1\right)-2e^{-2\kappa\tau}\left(2\kappa\tau+1\right)+2-e^{-3\kappa\tau}\right]\right\}.\\
\end{eqnarray*}
and

\begin{eqnarray*}
dD[M,\frac{1}{8}D[M,M]]_{t}&=& \frac{\nu^5\sigma_{t}}{8}dW_{t}\int^{T}_{t} e^{-\kappa(u-t)}\varphi(u)\left(\int_u^T e^{-\kappa(z-u)}\varphi(z)^2dz\right) du\\
&-& \frac{\nu^4}{8}\sigma^{2}_t\varphi(t)\left(\int_t^T e^{-\kappa(z-t)}\varphi(z)^2dz\right) dt
\end{eqnarray*}
\end{lemma}

\section{Proof of Theorem \ref{BS 2nd Approximation}}
The main idea employed in the proof of Theorem \ref{BS 2nd Approximation} is to apply Corollary \ref{BS Deco} to the call price formula iteratively. We also apply this corollary to the new terms that appear in the iterative procedure, and incorporate all the 
terms of order $O(\nu^{3})$ into the error term.

The term $(I)$ in formula (\ref{E:tri}) is decomposed using Corollary \ref{BS Deco} with $A_t=\Lambda\Gamma(BS)(t,X_{t},v_{t})$ and 
$B=U_{t}=\frac{\rho}{2}L[W,M]_{t}$. This gives

\begin{eqnarray}\label{GeneralExpansionI}
(I)&=&\frac{\rho}{2}\Lambda\Gamma(BS)(t,X_{t},v_{t}) L[W,M]_{t}\\ \nonumber
&+&\frac{\rho^{2}}{4}\mathbb{E}_{t}\left[\int^{T}_{t} e^{-r(u-t)} \Lambda^{2}\Gamma^{2}(BS)(u,X_{u},v_{u}) L[W,M]_{u} \sigma_{u} d[W,M]_{u}\right]\\ \nonumber
&+&\frac{\rho}{16}\mathbb{E}_{t}\left[\int^{T}_{t} e^{-r(u-t)}  \Lambda\Gamma^{3}(BS)(u,X_{u},v_{u}) L[W,M]_{u} d[M,M]_{u}\right]\\ \nonumber
&+&\frac{\rho^{2}}{2}\mathbb{E}_{t}\left[\int^{T}_{t} e^{-r(u-t)} \Lambda^{2}\Gamma(BS)(u,X_{u},v_{u})  \sigma_{u} d[W,L[W,M]]_{u}\right]\\ \nonumber
&+&\frac{\rho}{4}\mathbb{E}_{t}\left[\int^{T}_{t} e^{-r(u-t)} \Lambda\Gamma^{2}(BS)(u,X_{u},v_{u}) d[M,L[W,M]]_{u}\right].
\end{eqnarray}

Applying the same idea to the term $(II)$ in formula (\ref{E:tri}) and choosing $A_t=\Gamma^{2}(BS)(t,X_{t},v_{t}$ and 
$B=R_{t}=\frac{1}{8}D[M,M]_{t}$, we obtain

\begin{eqnarray}\label{GeneralExpansionII}
(II)&=&\frac{1}{8}\Gamma^{2}(BS)(t,X_{t},v_{t}) D[M,M]_{t}\\ \nonumber
&+&\frac{\rho}{16}\mathbb{E}_{t}\left[\int^{T}_{t} e^{-r(u-t)} \Lambda\Gamma^{3}(BS)(u,X_{u},v_{u}) D[M,M]_{u} \sigma_{u} d[W,M]_{u}\right]\\ \nonumber
&+&\frac{1}{64}\mathbb{E}_{t}\left[\int^{T}_{t} e^{-r(u-t)} \Gamma^{4}(BS)(u,X_{u},v_{u}) D[M,M]_{u} d[M,M]_{u}\right]\\ \nonumber
&+&\frac{\rho}{8}\mathbb{E}_{t}\left[\int^{T}_{t} e^{-r(u-t)} \Lambda \Gamma^{2}(BS)(u,X_{u},v_{u})  \sigma_{u} d[W,D[M,M]]_{u}\right]\\ \nonumber
&+&\frac{1}{16}\mathbb{E}_{t}\left[\int^{T}_{t} e^{-r(u-t)} \Gamma^{3}(BS)(u,X_{u},v_{u}) d[M,D[M,M]]_{u}\right].
\end{eqnarray}
In particular, using formula (\ref{GeneralExpansionI}) in the case of the Heston model, we find 
\begin{eqnarray*}
(I)&=&\frac{\rho\nu}{2}\Lambda \Gamma (BS)(t,X_{t},v_{t})\left(\int^{T}_{t}\mathbb{E}_{t} \left(\sigma_{s}^{2}\right)  \varphi(s) ds\right)\\
&+&\frac{\rho\nu^{3}}{16}\mathbb{E}_{t}\left[\int^{T}_{t}e^{-r(u-t)}\Lambda \Gamma^{3} (BS)(u,X_{u},v_{u})\left(\int^{T}_{u}\mathbb{E}_{u} \left(\sigma_{s}^{2}\right)  \varphi(s) ds\right)\sigma^{2}_{u}\varphi(u)^{2} du\right]\\ 
&+&\frac{\rho^{2}\nu^{2}}{4}\mathbb{E}_{t}\left[\int^{T}_{t}e^{-r(u-t)}\Lambda^{2} \Gamma^{2} (BS)(u,X_{u},v_{u})\left(\int^{T}_{u}\mathbb{E}_{u} \left(\sigma_{s}^{2}\right)  \varphi(s) ds\right)\sigma^{2}_{u}\varphi(u) du\right]\\ 
&+&\frac{\rho^{2} \nu^2}{2}\mathbb{E}_{t}\left[\int^{T}_{t}e^{-r(u-t)}\Lambda^{2} \Gamma (BS)(u,X_{u},v_{u})\left(\int^{T}_{u} e^{-\kappa(z-u)}\varphi(z)dz\right)\sigma^{2}_{u} du\right]\\ 
&+&\frac{\rho \nu^3}{4}\mathbb{E}_{t}\left[\int^{T}_{t}e^{-r(u-t)}\Lambda \Gamma^{2} (BS)(u,X_{u},v_{u})\left(\int^{T}_{u} e^{-\kappa(z-u)}\varphi(z)dz\right)\sigma^{2}_u\varphi(u) du\right]\\
&=&\Lambda \Gamma (BS)(t,X_{t},v_{t})U_{t}+(I.I) + (I.II) + (I.III) + (I.IV).
\end{eqnarray*}
Moreover, by applying formula (\ref{GeneralExpansionII}), we obtain 
\begin{eqnarray*}
(II)&=&\Gamma^{2} (BS)(t,X_{t},v_{t})R_{t}\\ 
&+&\frac{\nu^{4}}{64}\mathbb{E}_{t}\left[\int^{T}_{t}e^{-r(u-t)}\Gamma^{4} (BS)(u,X_{u},v_{u})\left(\int^{T}_{u}\mathbb{E}_{u}\left(\sigma^2_s\right) \varphi(s)^{2} ds\right)\sigma^{2}_{u}\varphi^{2}(u) du\right]\\ 
&+&\frac{\rho\nu^{3}}{16}\mathbb{E}_{t}\left[\int^{T}_{t}e^{-r(u-t)}\Lambda \Gamma^{3} (BS)(u,X_{u},v_{u})\left(\int^{T}_{u}\mathbb{E}_{u}\left(\sigma^2_s\right) \varphi(s)^{2} ds\right)\sigma^{2}_{u}\varphi(u)du\right]\\ 
&+&\frac{\rho\nu^3}{8}\mathbb{E}_{t}\left[\int^{T}_{t}e^{-r(u-t)}\Lambda\Gamma^{2} (BS)(u,X_{u},v_{u})\left(\int^{T}_{u} e^{-\kappa(z-u)}\varphi(z)^2dz\right) \sigma^{2}_{u} du\right]\\ 
&+&\frac{\nu^4}{16}\mathbb{E}_{t}\left[\int^{T}_{t}e^{-r(u-t)}\Gamma^{3} (BS)(u,X_{u},v_{u})\left(\int^{T}_{u} e^{-\kappa(z-u)}\varphi(z)^2dz\right) \varphi(u)\sigma^{2}_{u}du\right]\\
&=&\Gamma^{2} (BS)(t,X_{t},v_{t})R_{t} + (II.I) + (II.II) + (II.III) + (II.IV)
\end{eqnarray*}

\subsection{An upper bound for the term (I)}
The idea now is to discard terms of order $O(\nu^3)$, and apply Corollary \ref{BS Deco} to the terms of smaller order. 

\subsubsection{Upper bounds for the terms (I.I) and (I.IV)}
First, we note that the terms (I.I) and (I.IV) are of order $O(\nu^{3})$. Therefore, those terms can be incorporated into the error term. We have 

\begin{eqnarray*}
(I.I)+ (I.IV)&=&\frac{\rho\nu^{3}}{16}\mathbb{E}_{t}\left[\int^{T}_{t}e^{-r(u-t)}\Lambda \Gamma^{3}(BS)(u,X_{u},v_{u})\left(\int^{T}_{u}\mathbb{E}_{u} \left(\sigma_{s}^{2}\right)  \varphi(s) ds\right)\sigma^{2}_{u}\varphi(u)^{2} du\right]\\ 
&+&\frac{\rho \nu^3}{4}\mathbb{E}_{t}\left[\int^{T}_{t}e^{-r(u-t)}\Lambda \Gamma^{2} (BS)(u,X_{u},v_{u})\left(\int^{T}_{u} e^{-\kappa(z-u)}\varphi(z)dz\right)\sigma^{2}_u\varphi(u) du\right]\\
\end{eqnarray*}

Next, using Lemma \ref{fitage}, setting $a_{u}:=v_{u}\sqrt{T-u}$, and using the fact that $\varphi(t)$ is a decreasing function, we obtain

\begin{eqnarray*}
&&\left|(I.I)+ (I.IV)\right|\\
&\leq& C\frac{\rho\nu^{3}}{16}\mathbb{E}_{t}\left[\int^{T}_{t}e^{-r(u-t)}\left(\frac{1}{a^{6}_{u}}+ \frac{2}{a^{5}_{u}}+ \frac{1}{a^{4}_{u}}\right) \left(\int^{T}_{u}\mathbb{E}_{u} \left(\sigma_{s}^{2}\right) ds\right)\sigma^{2}_{u}\varphi(u)^{3} du\right]\\ 
&+&C\frac{\rho \nu^3}{4}\mathbb{E}_{t}\left[\int^{T}_{t}e^{-r(u-t)}\left(\frac{1}{a^{4}_{u}}
+ \frac{1}{a^{3}_{u}}\right)\sigma^{2}_u\varphi^{3}(u) du\right].\\
\end{eqnarray*}
It follows from Lemma \ref{tercerlema} (ii) that $\sigma^{2}_{t}\leq \frac{a^{2}_{t}}{\varphi(t)}$. Hence

\begin{eqnarray*}
\left|(I.I)+ (I.IV)\right|&\leq&C\frac{\rho\nu^{3}}{16}\mathbb{E}_{t}\left[\int^{T}_{t}e^{-r(u-t)}\left(\frac{5}{a^{2}_{u}}+ \frac{6}{a_{u}}+ 1\right) \varphi(u)^{2} du\right].\\
\end{eqnarray*}
Now, Lemma \ref{tercerlema} (i) implies that $a_{t}\geq\frac{\sqrt{\theta \kappa}}{\sqrt{2}}\varphi(t)$ and 

\begin{eqnarray*}
\left|(I.I)+ (I.IV)\right|&\leq&C\frac{\rho\nu^{3}}{16}\left[\int^{T}_{t}e^{-r(u-t)}\left(\frac{10}{\theta \kappa}+ \frac{6\sqrt{2}\varphi(u)}{\sqrt{\theta \kappa}}+ \varphi(u)^{2}\right) du\right]\\
\end{eqnarray*}
Finally, using the estimate $\varphi(t)\leq \frac{1}{\kappa}$, we see that
\begin{eqnarray*}
\left|(I.I)+ (I.IV)\right|&\leq&C\frac{\rho\nu^{3}}{16}\left(\frac{10}{\theta \kappa}+ \frac{6\sqrt{2}}{\kappa\sqrt{\theta \kappa}}+ \frac{1}{\kappa^{2}}\right) \left[\int^{T}_{t}e^{-r(u-t)}du\right].
\end{eqnarray*}

\subsubsection{An upper bound for the term (I.II)}
Here we note that the term (I.II) is of the order $O(\nu^{2})$. Therefore, in order to improve the approximation, we should apply 
Corollary \ref{BS Deco} to this term. Choosing $A_t=\frac{\rho^{2}\nu^{2}}{4}\Lambda^{2} \Gamma^{2} (BS)(t,X_{t},v_{t})$ and 
$B=\frac{1}{2}\left(\frac{\rho}{2}L[W,M]_{t}\right)^{2}$, we obtain

\begin{eqnarray*}
&&\left|(I.II)-\frac{1}{2}\Lambda^{2} \Gamma^{2} (BS)(t,X_{t},v_{t}) \left(\frac{\rho}{2}L[W,M]_{t}\right)^{2}\right|\\
&=&\frac{\nu\rho^{3}}{8}\mathbb{E}_{t}\left[\int^{T}_{t} e^{-r(u-t)}  \Lambda^{3} \Gamma^{3} (BS)(u,X_{u},v_{u}) L^{2}[W,M]_{u}\sigma^{2}_{u}\varphi(u)du\right]\\
&+&\frac{\rho^{2}\nu^{2}}{16}\mathbb{E}_{t}\left[\int^{T}_{t} e^{-r(u-t)}  \Lambda^{2} \Gamma^{4} (BS)(u,X_{u},v_{u}) L^{2}[W,M]_{u} \sigma^{2}_u\varphi^{2}(u)du\right]\\
&+&\frac{\rho^{3}\nu^2}{8}\mathbb{E}_{t}\left[\int^{T}_{t} e^{-r(u-t)} \Lambda^{3} \Gamma^{2} (BS)(u,X_{u},v_{u}) L[W,M]_{u} \sigma^{2}_{u} \left(\int_u^T e^{-\kappa(z-u)}\varphi(z)dz\right) du\right]\\
&+&\frac{\nu^3\rho^{2}}{8}\mathbb{E}_{t}\left[\int^{T}_{t} e^{-r(u-t)}  \Lambda^{2} \Gamma^{3} (BS)(u,X_{u},v_{u}) L[W,M]_{u} \sigma^{2}_u\varphi(u)\left(\int_u^T e^{-\kappa(z-u)}\varphi(z)dz\right) du\right]\\
&+&\frac{\rho^{2}\nu^4}{8}\mathbb{E}_{t}\left[\int^{T}_{t} e^{-r(u-t)}  \Lambda^{2} \Gamma^{2} (BS)(u,X_{u},v_{u})\left(\int_u^T e^{-\kappa(z-u)}\varphi(z)dz\right)^{2}\sigma^{2}_u du\right]\\
\end{eqnarray*}
Note that $L[W,M]_{t}\leq\nu a^{2}_{t}\varphi(t)$. It follows that
\begin{eqnarray*}
&&\left|(I.II)-\frac{1}{2}\Lambda^{2} \Gamma^{2} (BS)(t,X_{t},v_{t}) \left(\frac{\rho}{2}L[W,M]_{t}\right)^{2}\right|\\
&\leq&\frac{\nu^{3}\rho^{3}}{8}\mathbb{E}_{t}\left[\int^{T}_{t} e^{-r(u-t)} \left(\partial^{7}_{x}-2\partial^{6}_{x}+\partial^{5}_{x}\right)  \Gamma (BS)(u,X_{u},v_{u})  a^{4}_{u}\varphi^{3}(u)\sigma^{2}_{u}du\right]\\
&+&\frac{\rho^{2}\nu^{4}}{16}\mathbb{E}_{t}\left[\int^{T}_{t} e^{-r(u-t)}\left(\partial^{8}_{x}-3\partial^{7}_{x}+3\partial^{6}_{x}-\partial^{5}_{x}\right) \Gamma (BS)(u,X_{u},v_{u}) a^{4}_{u}\varphi^{4}(u) \sigma^{2}_u du\right]\\
&+&\frac{\rho^{3}\nu^3}{8}\mathbb{E}_{t}\left[\int^{T}_{t} e^{-r(u-t)} \left(\partial^{5}_{x}-\partial^{4}_{x}\right) \Gamma (BS)(u,X_{u},v_{u})  a^{2}_{u}\varphi^{3}(u) \sigma^{2}_{u} du\right]\\
&+&\frac{\nu^4\rho^{2}}{8}\mathbb{E}_{t}\left[\int^{T}_{t} e^{-r(u-t)}  \left(\partial^{6}_{x}-2\partial^{5}_{x}+\partial^{4}_{x}\right) \Gamma (BS)(u,X_{u},v_{u}) a^{2}_{u}\varphi^{4}(u) \sigma^{2}_u du\right]\\
&+&\frac{\rho^{2}\nu^4}{8}\mathbb{E}_{t}\left[\int^{T}_{t} e^{-r(u-t)}  \left(\partial^{4}_{x}-\partial^{3}_{x}\right) \Gamma (BS)(u,X_{u},v_{u}) \varphi^{4}(u)\sigma^{2}_u du\right]\\
\end{eqnarray*}
Next, using Lemma \ref{fitage}, we get
\begin{eqnarray*}
&&\left|(I.II)-\frac{1}{2}\Lambda^{2} \Gamma^{2} (BS)(t,X_{t},v_{t}) \left(\frac{\rho}{2}L[W,M]_{t}\right)^{2}\right|\\
&\leq&\frac{C\nu^{3}\rho^{3}}{8}\mathbb{E}_{t}\left[\int^{T}_{t} e^{-r(u-t)} \left(\frac{1}{a^{4}_{u}}+\frac{2}{a^{3}_{u}}+\frac{1}{a^{2}_{u}}\right)  \varphi^{3}(u)\sigma^{2}_{u}du\right]\\
&+&\frac{C\rho^{2}\nu^{4}}{16}\mathbb{E}_{t}\left[\int^{T}_{t} e^{-r(u-t)}\left(\frac{1}{a^{5}_{u}}+\frac{3}{a^{4}_{u}}+\frac{3}{a^{3}_{u}}+\frac{1}{a^{2}_{u}}\right)  \varphi^{4}(u) \sigma^{2}_u du\right]\\
&+&\frac{C\rho^{3}\nu^3}{8}\mathbb{E}_{t}\left[\int^{T}_{t} e^{-r(u-t)} \left(\frac{1}{a^{4}_{u}}+\frac{1}{a^{3}_{u}}\right)   \varphi^{3}(u) \sigma^{2}_{u} du\right]\\
&+&\frac{C\nu^4\rho^{2}}{8}\mathbb{E}_{t}\left[\int^{T}_{t} e^{-r(u-t)}  \left(\frac{1}{a^{5}_{u}}+\frac{2}{a^{4}_{u}}+\frac{1}{a^{3}_{u}}\right) \varphi^{4}(u) \sigma^{2}_u du\right]\\
&+&\frac{C\rho^{2}\nu^4}{128}\mathbb{E}_{t}\left[\int^{T}_{t} e^{-r(u-t)}  \left(\frac{1}{a^{5}_{u}}+\frac{1}{a^{4}_{u}}\right) \varphi^{4}(u)\sigma^{2}_u du\right].
\end{eqnarray*}
Now, Lemma \ref{tercerlema} (ii) implies that $\sigma^{2}_{t}\leq \frac{a^{2}_{t}}{\varphi(t)}$ and 

\begin{eqnarray*}
&&\left|(I.II)-\frac{1}{2}\Lambda^{2} \Gamma^{2} (BS)(t,X_{t},v_{t}) \left(\frac{\rho}{2}L[W,M]_{t}\right)^{2}\right|\\
&\leq&\frac{C\nu^{3}\rho^{3}}{8}\mathbb{E}_{t}\left[\int^{T}_{t} e^{-r(u-t)} \left(\frac{2}{a^{2}_{u}}+\frac{3}{a_{u}}+1\right)  \varphi^{2}(u) du\right]\\
&+&\frac{C\rho^{2}\nu^{4}}{16}\mathbb{E}_{t}\left[\int^{T}_{t} e^{-r(u-t)}\left(\frac{5}{a^{3}_{u}}+\frac{9}{a^{2}_{u}}+\frac{3}{a_{u}}+1\right)  \varphi^{3}(u) du\right].
\end{eqnarray*}
Moreover, Lemma \ref{tercerlema} gives $a_{t}\geq\frac{\sqrt{\theta \kappa}}{\sqrt{2}}\varphi(t)$ and 

\begin{eqnarray*}
&&\left|(I.II)-\frac{1}{2}\Lambda^{2} \Gamma^{2} (BS)(t,X_{t},v_{t}) \left(\frac{\rho}{2}L[W,M]_{t}\right)^{2}\right|\\
&\leq&\frac{C\nu^{3}\rho^{3}}{8}\left[\int^{T}_{t} e^{-r(u-t)} \left(\frac{8}{\theta \kappa }+\frac{3\sqrt{2}\varphi(u)}{\sqrt{\theta \kappa}}+\varphi^{2}(u)\right)   du\right]\\
&+&\frac{C\rho^{2}\nu^{4}}{16}\left[\int^{T}_{t} e^{-r(u-t)}\left(\frac{10\sqrt{2}}{\theta \kappa\sqrt{\theta \kappa}}+\frac{18\varphi^{2}(u)}{\theta \kappa }+\frac{3\sqrt{2}\varphi^{2}(u)}{\sqrt{\theta \kappa}}+\varphi^{3}(u)\right)   du\right].
\end{eqnarray*}

Finally, using the estimate $\varphi(t)\leq \frac{1}{\kappa}$, we obtain

\begin{eqnarray*}
&&\left|(I.II)-\frac{1}{2}\Lambda^{2} \Gamma^{2} (BS)(t,X_{t},v_{t}) \left(\frac{\rho}{2}L[W,M]_{t}\right)^{2}\right|\\
&\leq&\frac{C\nu^{3}\rho^{3}}{8}\left(\frac{8}{\theta \kappa }+\frac{3\sqrt{2}}{\kappa\sqrt{\theta \kappa}}+\frac{1}{\kappa^{2}}\right)\left[\int^{T}_{t} e^{-r(u-t)}    du\right]\\
&+&\frac{C\rho^{2}\nu^{4}}{16}\left(\frac{10\sqrt{2}}{\theta \kappa\sqrt{\theta \kappa}}+\frac{18}{\theta \kappa^{3} }+\frac{3\sqrt{2}}{\kappa^{3}\sqrt{\theta \kappa}}+\frac{1}{\kappa^{3}}\right)\left[\int^{T}_{t} e^{-r(u-t)}   du\right]\\
\end{eqnarray*}

\subsubsection{An upper bound for the term (I.III)}
The term (I.III) is of order $O(\nu^{2})$, and it has to be taken into account in the approximation. Here we apply Corollary \ref{BS Deco} with $A_t=\Lambda^{2} \Gamma (BS)(t,X_{t},v_{t})$ and $B=\rho L[W,\frac{\rho}{2}L[W,M]]_{t}$.
We have

\begin{eqnarray*}
&&\left|(I.III) - \rho\Lambda^{2} \Gamma (BS)(t,X_{t},v_{t}) L[W,\frac{\rho}{2}L[W,M]]_{t}\right|\\
&=&\frac{\rho^{3}\nu}{4}\mathbb{E}_{t}\left[\int^{T}_{t} e^{-r(u-t)} \Lambda^{3} \Gamma^{2} (BS)(u,X_{u},v_{u})  L[W,L[W,M]]_{u} \sigma^{2}_{u} \varphi(u) du\right]\\
&+&\frac{\rho^{2}\nu^{2}}{16}\mathbb{E}_{t}\left[\int^{T}_{t} e^{-r(u-t)} \Lambda^{2} \Gamma^{3} (BS)(u,X_{u},v_{u}) L[W,L[W,M]]_{u} \sigma^{2}_{u} \varphi^{2}(u)du\right]\\
&+&\frac{\rho^{3}\nu^3}{4}\mathbb{E}_{t}\left[\int^{T}_{t} e^{-r(u-t)}  \Lambda^{3} \Gamma (BS)(u,X_{u},v_{u})  \sigma^{2}_{u} \left[\int^{T}_{u}\left(\int^{T}_{s} e^{-\kappa(z-s)}\varphi(z)dz\right)e^{-\kappa(s-u)} ds\right]du\right]\\
&+&\frac{\rho^{2}\nu^4}{8}\mathbb{E}_{t}\left[\int^{T}_{t} e^{-r(u-t)} \Lambda^{2} \Gamma^{2} (BS)(u,X_{u},v_{u}) \sigma^{2}_u\varphi(u)\left[\int^{T}_{u}\left(\int^{T}_{s} e^{-\kappa(z-s)}\varphi(z)dz\right)e^{-\kappa(s-u)} ds\right]du\right]\\
&+&\frac{\rho^{4} \nu^6}{16}\mathbb{E}_{t}\left[\int^{T}_{t} e^{-r(u-t)} \Lambda^{2} \Gamma (BS)(u,X_{u},v_{u}) \left[\int^{T}_{u}\left(\int^{T}_{s} e^{-\kappa(z-s)}\varphi(z)dz\right)e^{-\kappa(s-u)} ds\right]^{2}\sigma^{2}_{u} du \right].\\
\end{eqnarray*}
It is easy to see that $L[W,L[W,M]]_{t}=\nu^2  a^{2}_{t} \varphi^{2}(t)$. It follows that

\begin{eqnarray*}
&&\left|(I.III) - \rho\Lambda^{2} \Gamma (BS)(t,X_{t},v_{t}) L[W,\frac{\rho}{2}L[W,M]]_{t}\right|\\
&\leq&\frac{\rho^{3}\nu^3}{4}\mathbb{E}_{t}\left[\int^{T}_{t} e^{-r(u-t)} \left(\partial^{5}_{x}-\partial^{3}_{x}\right) \Gamma (BS)(u,X_{u},v_{u})   a^{2}_{u} \varphi^{3}(u) \sigma^{2}_{u}  du\right]\\
&+&\frac{\rho^{2}\nu^{4}}{16}\mathbb{E}_{t}\left[\int^{T}_{t} e^{-r(u-t)} \left(\partial^{6}_{x}-2\partial^{5}_{x}+\partial^{4}_{x}\right) \Gamma (BS)(u,X_{u},v_{u})  a^{2}_{u} \varphi^{4}(u)  \sigma^{2}_{u} du\right]\\
&+&\frac{\rho^{3}\nu^3}{4}\mathbb{E}_{t}\left[\int^{T}_{t} e^{-r(u-t)}  \partial^{3}_{x} \Gamma (BS)(u,X_{u},v_{u})  \sigma^{2}_{u} \varphi^{3}(u)du\right]\\
&+&\frac{\rho^{2}\nu^4}{8}\mathbb{E}_{t}\left[\int^{T}_{t} e^{-r(u-t)} \left(\partial^{4}_{x}-\partial^{3}_{x}\right) \Gamma (BS)(u,X_{u},v_{u}) \sigma^{2}_u\varphi^{4}(u)du\right]\\
&+&\frac{\rho^{4} \nu^6}{16}\mathbb{E}_{t}\left[\int^{T}_{t} e^{-r(u-t)} \partial^{2}_{x} \Gamma (BS)(u,X_{u},v_{u}) \varphi^{6}(u)\sigma^{2}_{u} du \right].
\end{eqnarray*}
Nest, using Lemma \ref{fitage}, we get

\begin{eqnarray*}
&&\left|(I.III) - \rho\Lambda^{2} \Gamma (BS)(t,X_{t},v_{t}) L[W,\frac{\rho}{2}L[W,M]]_{t}\right|\\
&\leq&\frac{C\rho^{3}\nu^3}{4}\mathbb{E}_{t}\left[\int^{T}_{t} e^{-r(u-t)} \left(\frac{1}{a_{u}^{4}}+\frac{1}{a_{u}^{2}}\right)    \varphi^{3}(u) \sigma^{2}_{u}  du\right]\\
&+&\frac{C\rho^{2}\nu^{4}}{16}\mathbb{E}_{t}\left[\int^{T}_{t} e^{-r(u-t)} \left(\frac{1}{a_{u}^{5}}+2\frac{1}{a_{u}^{4}}+\frac{1}{a_{u}^{3}}\right)  \varphi^{4}(u)  \sigma^{2}_{u} du\right]\\
&+&\frac{C\rho^{3}\nu^3}{4}\mathbb{E}_{t}\left[\int^{T}_{t} e^{-r(u-t)}  \frac{1}{a_{u}^{4}} \sigma^{2}_{u} \varphi^{3}(u)du\right]\\
&+&\frac{C\rho^{2}\nu^4}{8}\mathbb{E}_{t}\left[\int^{T}_{t} e^{-r(u-t)} \left(\frac{1}{a_{u}^{5}}+\frac{1}{a_{u}^{4}}\right)  \sigma^{2}_u\varphi^{4}(u)du\right]\\
&+&\frac{C\rho^{4} \nu^6}{16}\mathbb{E}_{t}\left[\int^{T}_{t} e^{-r(u-t)} \frac{1}{a_{u}^{3}}  \varphi^{6}(u)\sigma^{2}_{u} du \right].
\end{eqnarray*}
Now, Lemma \ref{tercerlema} (ii), gives $\sigma^{2}_{t}\leq \frac{a^{2}_{t}}{\varphi(t)}$ and 

\begin{eqnarray*}
&&\left|(I.III) - \rho\Lambda^{2} \Gamma (BS)(t,X_{t},v_{t}) L[W,\frac{\rho}{2}L[W,M]]_{t}\right|\\
&\leq&\frac{C\rho^{3}\nu^3}{4}\mathbb{E}_{t}\left[\int^{T}_{t} e^{-r(u-t)} \left(\frac{1}{a_{u}^{2}}+1\right)    \varphi^{2}(u)   du\right]\\
&+&\frac{C\rho^{2}\nu^{4}}{16}\mathbb{E}_{t}\left[\int^{T}_{t} e^{-r(u-t)} \left(\frac{1}{a_{u}^{3}}+2\frac{1}{a_{u}^{2}}+\frac{1}{a_{u}^{1}}\right)  \varphi^{3}(u)    du\right]\\
&+&\frac{C\rho^{3}\nu^3}{4}\mathbb{E}_{t}\left[\int^{T}_{t} e^{-r(u-t)}  \frac{1}{a_{u}^{2}}   \varphi^{2}(u)du\right]\\
&+&\frac{C\rho^{2}\nu^4}{8}\mathbb{E}_{t}\left[\int^{T}_{t} e^{-r(u-t)} \left(\frac{1}{a_{u}^{3}}+\frac{1}{a_{u}^{2}}\right)  \varphi^{3}(u)du\right]\\
&+&\frac{C\rho^{4} \nu^6}{16}\mathbb{E}_{t}\left[\int^{T}_{t} e^{-r(u-t)} \frac{1}{a_{u}}  \varphi^{5}(u)  du \right]\\
\end{eqnarray*}
Moreover, Lemma \ref{tercerlema} (i) implies that that $a_{t}\geq\frac{\sqrt{\theta \kappa}}{\sqrt{2}}\varphi(t)$ and 

\begin{eqnarray*}
&&\left|(I.III) - \rho\Lambda^{2} \Gamma (BS)(t,X_{t},v_{t}) L[W,\frac{\rho}{2}L[W,M]]_{t}\right|\\
&\leq&\frac{C\rho^{3}\nu^3}{4}\left[\int^{T}_{t} e^{-r(u-t)} \left(\frac{2}{\theta \kappa \varphi(u)}+1\right)    \varphi^{2}(u)   du\right]\\
&+&\frac{C\rho^{2}\nu^{4}}{16}\left[\int^{T}_{t} e^{-r(u-t)} \left(\frac{2\sqrt{2}}{\theta \kappa\sqrt{\theta \kappa}\varphi^{2}(u)}+\frac{4}{\theta \kappa \varphi(u)}+\frac{\sqrt{2}}{\sqrt{\theta \kappa}}\right)  \varphi^{2}(u)    du\right]\\
&+&\frac{C\rho^{3}\nu^3}{4}\left[\int^{T}_{t} e^{-r(u-t)}  \frac{2}{\theta \kappa}   du\right]\\
&+&\frac{C\rho^{2}\nu^4}{8}\left[\int^{T}_{t} e^{-r(u-t)} \left(\frac{2\sqrt{2}}{\theta \kappa\sqrt{\theta \kappa}\varphi(u)}+\frac{2}{\theta \kappa}\right)  \varphi(u)du\right]\\
&+&\frac{C\rho^{4} \nu^6}{16}\left[\int^{T}_{t} e^{-r(u-t)} \frac{\sqrt{2}}{\sqrt{\theta \kappa}}  \varphi^{4}(u)  du \right]\\
\end{eqnarray*}
Next, using the estimate $\varphi(t)\leq \frac{1}{\kappa}$, we obtain

\begin{eqnarray*}
&&\left|(I.III) - \rho\Lambda^{2} \Gamma (BS)(t,X_{t},v_{t}) L[W,\frac{\rho}{2}L[W,M]]_{t}\right|\\
&\leq&\frac{C\rho^{3}\nu^3}{4}\left[\int^{T}_{t} e^{-r(u-t)} \left(\frac{2}{\theta }+1\right)   \frac{1}{\kappa^{2}}   du\right]\\
&+&\frac{C\rho^{2}\nu^{4}}{16}\left[\int^{T}_{t} e^{-r(u-t)} \left(\frac{2\sqrt{2\kappa}}{\theta \sqrt{\theta }}+\frac{4}{\theta }+\frac{\sqrt{2}}{\sqrt{\theta \kappa}}\right) \frac{1}{\kappa^{2}}    du\right]\\
&+&\frac{C\rho^{3}\nu^3}{4}\left[\int^{T}_{t} e^{-r(u-t)}  \frac{2}{\theta \kappa}   du\right]\\
&+&\frac{C\rho^{2}\nu^4}{8}\left[\int^{T}_{t} e^{-r(u-t)} \left(\frac{2\sqrt{2}}{\theta \sqrt{\theta \kappa}}+\frac{2}{\theta \kappa}\right)  \frac{1}{\kappa}du\right]\\
&+&\frac{C\rho^{4} \nu^6}{16}\left[\int^{T}_{t} e^{-r(u-t)} \frac{\sqrt{2}}{\sqrt{\theta \kappa}}  \frac{1}{\kappa^{4}}  du \right].\\
\end{eqnarray*}
Therefore 
\begin{eqnarray*}
&&\left|(I.III) - \rho\Lambda^{2} \Gamma (BS)(t,X_{t},v_{t}) L[W,\frac{\rho}{2}L[W,M]]_{t}\right|\\
&\leq&\frac{C\rho^{3}\nu^3}{4\kappa^{2}}\left(\frac{2\kappa+2}{\theta }+1\right) \left[\int^{T}_{t} e^{-r(u-t)}    du\right]\\
&+&\frac{C\rho^{2}\nu^{4}}{16\kappa^{2}}\left(\frac{6\sqrt{2\kappa}}{\theta \sqrt{\theta }}+\frac{8}{\theta }+\frac{\sqrt{2}}{\sqrt{\theta \kappa}}\right) \left[\int^{T}_{t} e^{-r(u-t)}     du\right]\\
&+&\frac{C\rho^{4} \nu^6}{16}\frac{\sqrt{2}}{\kappa^{4}\sqrt{\theta \kappa}}\left[\int^{T}_{t} e^{-r(u-t)} du \right].\\
\end{eqnarray*}

\subsection{An upper bound for the term (II)}
In this case, all the terms can be incorporated into the error term. Using the fact that $\varphi(t)$ is a decreasing function, we obtain
\begin{eqnarray*}
&&\left|(II)-\Gamma^{2}(BS)(t,X_{t},v_{t})R_{t}\right|\\ 
&\leq&\frac{\nu^{4}}{64}\mathbb{E}_{t}\left[\int^{T}_{t}e^{-r(u-t)}\left(\partial^{6}_{x}-3\partial^{5}_{x} + 3\partial^{4}_{x} - \partial^{3}_{x}\right)\Gamma BS(u,X_{u},v_{u})a^{2}_{u}\sigma^{2}_{u}\varphi^{4}(u) du\right]\\ 
&+&\frac{\rho\nu^{3}}{16}\mathbb{E}_{t}\left[\int^{T}_{t}e^{-r(u-t)}\left(\partial^{5}_{x}-2\partial^{4}_{x} + \partial^{3}_{x}\right) \Gamma BS(u,X_{u},v_{u})a^{2}_{u}\sigma^{2}_{u}\varphi^{3}(u)du\right]\\ 
&+&\frac{\rho\nu^3}{8}\mathbb{E}_{t}\left[\int^{T}_{t}e^{-r(u-t)}\left(\partial^{3}_{x}-\partial^{2}_{x}\right)\Gamma BS(u,X_{u},v_{u}) \sigma^{2}_{u} \varphi(u)^3 du\right]\\ 
&+&\frac{\nu^4}{16}\mathbb{E}_{t}\left[\int^{T}_{t}e^{-r(u-t)}\left(\partial^{4}_{x}-2\partial^{3}_{x} + \partial^{2}_{x}\right)\Gamma BS(u,X_{u},v_{u}) \varphi^{4}(u)\sigma^{2}_{u}du\right].\\
\end{eqnarray*}
Next, using Lemma \ref{fitage} and setting $a_{u}:=v_{u}\sqrt{T-u}$, we get
\begin{eqnarray*}
&&\left|(II)-\Gamma^{2} (BS)(t,X_{t},v_{t})R_{t}\right|\\ 
&\leq&\frac{C\nu^{4}}{64}\mathbb{E}_{t}\left[\int^{T}_{t}e^{-r(u-t)}\left(\frac{1}{a^{5}_{u}}+\frac{3}{a^{4}_{u}} + \frac{3}{a^{3}_{u}} + \frac{1}{a^{2}_{u}}\right)\sigma^{2}_{u}\varphi^{4}(u) du\right]\\ 
&+&\frac{C\rho\nu^{3}}{16}\mathbb{E}_{t}\left[\int^{T}_{t}e^{-r(u-t)}\left(\frac{1}{a^{4}_{u}}+\frac{2}{a^{3}_{u}} + \frac{1}{a^{2}_{u}}\right) \sigma^{2}_{u}\varphi^{3}(u)du\right]\\ 
&+&\frac{C\rho\nu^3}{8}\mathbb{E}_{t}\left[\int^{T}_{t}e^{-r(u-t)}\left(\frac{1}{a^{4}_{u}}+\frac{1}{a^{3}_{u}}\right) \sigma^{2}_{u} \varphi(u)^3 du\right]\\ 
&+&\frac{C\nu^4}{16}\mathbb{E}_{t}\left[\int^{T}_{t}e^{-r(u-t)}\left(\frac{1}{a^{5}_{u}}+\frac{2}{a^{4}_{u}} + \frac{1}{a^{3}_{u}}\right) \varphi^{4}(u)\sigma^{2}_{u}du\right]\\
\end{eqnarray*}
It follows from Lemma \ref{tercerlema} (ii) that $\sigma^{2}_{t}\leq \frac{a^{2}_{t}}{\varphi(t)}$ and 

\begin{eqnarray*}
&&\left|(II)-\Gamma^{2} (BS)(t,X_{t},v_{t})R_{t}\right|\\ 
&\leq&\frac{C\nu^{4}}{64}\mathbb{E}_{t}\left[\int^{T}_{t}e^{-r(u-t)}\left(\frac{1}{a^{3}_{u}}+\frac{3}{a^{2}_{u}} + \frac{3}{a_{u}} + 1\right)\varphi^{3}(u) du\right]\\ 
&+&\frac{C\rho\nu^{3}}{16}\mathbb{E}_{t}\left[\int^{T}_{t}e^{-r(u-t)}\left(\frac{1}{a^{2}_{u}}+\frac{2}{a_{u}} + 1\right) \varphi^{2}(u)du\right]\\ 
&+&\frac{C\rho\nu^3}{8}\mathbb{E}_{t}\left[\int^{T}_{t}e^{-r(u-t)}\left(\frac{1}{a^{2}_{u}}+\frac{1}{a_{u}}\right)  \varphi^2(u) du\right]\\ 
&+&\frac{C\nu^4}{16}\mathbb{E}_{t}\left[\int^{T}_{t}e^{-r(u-t)}\left(\frac{1}{a^{3}_{u}}+\frac{2}{a^{2}_{u}} + \frac{1}{a_{u}}\right) \varphi^{3}(u)du\right].\\
\end{eqnarray*}
Therefore, Lemma \ref{tercerlema} (i) gives $a_{t}\geq\frac{\sqrt{\theta \kappa}}{\sqrt{2}}\varphi(t)$ and 

\begin{eqnarray*}
&&\left|(II)-\Gamma^{2} (BS)(t,X_{t},v_{t})R_{t}\right|\\ 
&\leq&\frac{C\nu^{4}}{64}\left[\int^{T}_{t}e^{-r(u-t)}\left(\frac{2\sqrt{2}}{\theta \kappa\sqrt{\theta \kappa}\varphi^{3}(u)}+\frac{6}{\theta \kappa\varphi^{2}(u)} + \frac{3\sqrt{2}}{\sqrt{\theta \kappa} \varphi(u)} + 1\right)\varphi^{3}(u) du\right]\\ 
&+&\frac{C\rho\nu^{3}}{16}\left[\int^{T}_{t}e^{-r(u-t)}\left(\frac{2}{\theta \kappa \varphi^{2}(u)}+\frac{2\sqrt{2}}{\sqrt{\theta \kappa}\varphi(u)} + 1\right) \varphi^{2}(u)du\right]\\ 
&+&\frac{C\rho\nu^3}{8}\left[\int^{T}_{t}e^{-r(u-t)}\left(\frac{2}{\theta \kappa \varphi(u)}+\frac{\sqrt{2}}{\sqrt{\theta \kappa}}\right)  \varphi(u) du\right]\\ 
&+&\frac{C\nu^4}{16}\left[\int^{T}_{t}e^{-r(u-t)}\left(\frac{2\sqrt{2}}{\theta \kappa\sqrt{\theta \kappa}\varphi^{2}(u)}+\frac{4}{\theta \kappa \varphi(u)} + \frac{\sqrt{2}}{\sqrt{\theta \kappa}}\right) \varphi^{2}(u)du\right].\\
\end{eqnarray*}

Finally, we observe that the estimate $\varphi(t)\leq \frac{1}{\kappa}$ implies that

\begin{eqnarray*}
&&\left|(II)-\Gamma^{2} (BS)(t,X_{t},v_{t})R_{t}\right|\\ 
&\leq&\frac{C\nu^{4}}{64\kappa^{3}}\left(\frac{2\kappa\sqrt{2\kappa}(1+4\kappa)}{\theta \sqrt{\theta}}+\frac{22\kappa}{\theta } + \frac{7\sqrt{2\kappa}}{\sqrt{\theta } } + 1\right)\left[\int^{T}_{t}e^{-r(u-t)} du\right]\\ 
&+&\frac{C\rho\nu^{3}}{16\kappa^{2}}\left(\frac{6\kappa + 2\sqrt{2\kappa}}{\theta }+\frac{2\sqrt{2\kappa}}{\sqrt{\theta}} + 1\right) 
\left[\int^{T}_{t}e^{-r(u-t)}du\right].\\ 
\end{eqnarray*}

This completes the proof of Theorem \ref{BS 2nd Approximation}.

\section{Sketch of the proof of Theorem \ref{BS Approximation2}}
The proof of Theorem \ref{BS Approximation2} follows the same arguments as the previous proof. We will next provide a sketch of the proof and skip the lengthy computations. The main idea employed in the proof is to keep applying Corollary \ref{BS Deco} to all the terms with order lower than $O(\nu^{4})$, and to estimate the appearing new terms. Our next goal is to describe the terms that have to be decomposed. 

The term (I.I) is decomposed into a series of new terms. We approximate (I.I) by the following expression:
\begin{eqnarray*}
A_t=\Lambda^{2}\Gamma^{2} BS(t,X_{t},v_{t}) \text{ and } B_t=\frac{1}{2}\left(\frac{\rho}{2}L[W,M]_{t}\right)^{2}.
\end{eqnarray*}
This gives
\begin{eqnarray*}
&-&\frac{\rho^{2}}{4}\mathbb{E}_{t}\left[\int^{T}_{t} e^{-r(u-t)} \Lambda^{2}\Gamma^{2} BS_{u} L[W,M]_{u}\sigma_{u}d[W,M]_{u}\right]\\
&=&\frac{1}{2}\Lambda^{2}\Gamma^{2} BS_{t} \left(\frac{\rho}{2}L[W,M]_{t}\right)^{2}\\
&+&\frac{\rho^3}{32}\mathbb{E}_{t}\left[\int^{T}_{t} e^{-r(u-t)} \Lambda^{3}\Gamma^{3} BS_{u} L^{2}[W,M]_{u} \sigma_{u} d[W,M]_{u}\right]\\
&+&\frac{\rho^{2}}{128}\mathbb{E}_{t}\left[\int^{T}_{t} e^{-r(u-t)} \Lambda^{2}\Gamma^{4} BS_{u} L^{2}[W,M]_{u} d[M,M]_{u}\right]\\
&+&\frac{\rho^{4}}{32}\mathbb{E}_{t}\left[\int^{T}_{t} e^{-r(u-t)} \Lambda^{3}\Gamma^{2} BS_{u} L[W,M]_{u} \sigma_{u} d[W,L^{2}[W,M]]_{u}\right]\\
&+&\frac{\rho^{3}}{64}\mathbb{E}_{t}\left[\int^{T}_{t} e^{-r(u-t)} \Lambda^{2}\Gamma^{3} BS_{u} L[W,M]_{u} d[M,L^{2}[W,M]]_{u}\right]\\
&+&\frac{\rho^{4}}{128}\mathbb{E}_{t}\left[\int^{T}_{t} e^{-r(u-t)} \Lambda^{2}\Gamma^{2} BS_{u} d[L^{2}[W,M],L^{2}[W,M]]_{u}\right]\\ &=& \frac{1}{2}\Lambda^{2}\Gamma^{2} BS_{t} \left(\frac{\rho}{2}L[W,M]_{t}\right)^{2}+(I.I.I) + \ldots + (I.I.V).
\end{eqnarray*}

The terms (I.II) and (II.I) are approximated by 
\begin{eqnarray*}
A_t=\Lambda\Gamma^3 BS_t
\text{ and }
B_t=\left(\frac{\rho}{2}L[W,M]_{t}\right)\left(\frac{1}{8}D[M,M]_{t}\right),
\end{eqnarray*}
while the term (I.III) is approximated by 
\begin{eqnarray*}
A_t=\Lambda^{2}\Gamma BS(t,X_{t},v_{t})
\text{ and }
B_t=\frac{\rho^{2}}{2}L[W,L[W,M]]_{t}.
\end{eqnarray*}
It follows that
\begin{eqnarray*}
&-&\frac{\rho^{2}}{2}\mathbb{E}_{t}\left[\int^{T}_{t} e^{-r(u-t)} \Lambda^{2}\Gamma BS_{u} \sigma_{u} d[W,L[W,M]]_{u} \right]\\
&=&\frac{\rho^{2}}{2} \Lambda^{2}\Gamma BS_{t} L[W,L[W,M]]_{t}\\
&+&\frac{\rho^{3}}{4}\mathbb{E}_{t}\left[\int^{T}_{t} e^{-r(u-t)} \Lambda^{3}\Gamma^{2} BS_{u} L[W,L[W,M]]_{u} \sigma_{u} d[W,M]_{u}\right]\\
&+&\frac{\rho^{2}}{16}\mathbb{E}_{t}\left[\int^{T}_{t} e^{-r(u-t)} \Lambda^{2}\Gamma^{3} BS_{u} L[W,L[W,M]]_{u} d[M,M]_{u}\right]\\
&+&\frac{\rho^{3}}{2}\mathbb{E}_{t}\left[\int^{T}_{t} e^{-r(u-t)} \Lambda^{3}\Gamma BS_{u}  \sigma_{u} d[W,L[W,L[W,M]]]_{u}\right]\\
&+&\frac{\rho^{2}}{4}\mathbb{E}_{t}\left[\int^{T}_{t} e^{-r(u-t)} \Lambda^{2}\Gamma^{2} BS_{u} d[M,L[W,L[W,M]]]_{u}\right]\\
&=& \Lambda^{2}\Gamma BS_{t} \frac{\rho^{2}}{2}L[W,L[W,M]]_{t}+ (I.III.I) + \ldots (I.III.IV).
\end{eqnarray*}

The term (I.IV) is approximated by 
\begin{eqnarray*}
A_t=\Lambda\Gamma^{2} BS_t
\text{ and }
B_t=\frac{\rho}{4}D[M,L[W,M]]_{u},
\end{eqnarray*}
while the term (II.III) is approximated by 
\begin{eqnarray*}
A_t=\Lambda \Gamma^2 BS_t
\text{ and }
B_t=\rho L[W,\frac{1}{8}D[M,M]]_{t}.
\end{eqnarray*}
Similarly, the term (I.I.I) is approximated by 
\begin{eqnarray*}
A_t=\Lambda^{3}\Gamma^3 BS_t
\text{ and }
B_t=\frac{1}{6}\left(\frac{\rho}{2}L[W,M]_{t}\right)^{3}.
\end{eqnarray*}

The terms (I.I.III) and (I.III.I) are approximated by the following expressions:
\begin{eqnarray*}
A_t=\Lambda^{3}\Gamma^2 BS_t
\text{ and }
B_t=\frac{\rho^{3}}{4}L[W,M]_{t}L[W,L[W,M]]_{t}.
\end{eqnarray*}
Moreover, the term (I.III.III) is approximated by 
\begin{eqnarray*}
A_t=\Lambda^{3}\Gamma BS_t
\text{ and }
B_t=\frac{\rho^{3}}{2}L[W,L[W,L[W,M]]]_{t}. 
\end{eqnarray*}

We estimate each of the new terms appearing in the proof exactly as in the previous proof.

\section{Proof of Theorem \ref{BS Approximation zero-corr}}
It is easy to see that for the uncorrelated Heston model we have
\begin{eqnarray*}
V_{t}&=&(BS)(t,X_{t},v_{t}) \nonumber \\
&+&\frac{1}{8}\mathbb{E}_{t}\left[\int^{T}_{t}e^{-r(u-t)}\Gamma^{2}(BS)(u,X_{u},v_{u})d[M,M]_{u}\right]. \nonumber
\end{eqnarray*}
Using Corollary \ref{BS Deco}, we obtain a special case of formula (\ref{GeneralExpansionII}). In particular, we have

\begin{eqnarray*}
(II)&=&\frac{1}{8}\Gamma^{2}(BS)(t,X_{t},v_{t}) D[M,M]_{t}\\ \nonumber
&+&\frac{1}{64}\mathbb{E}_{t}\left[\int^{T}_{t} e^{-r(u-t)} \Gamma^{4}(BS)(u,X_{u},v_{u}) D[M,M]_{u} d[M,M]_{u}\right]\\ \nonumber
&+&\frac{1}{16}\mathbb{E}_{t}\left[\int^{T}_{t} e^{-r(u-t)} \Gamma^{3}(BS)(u,X_{u},v_{u}) d[M,D[M,M]]_{u}\right]\\ \nonumber
&=&(A)+(B).
\end{eqnarray*}

\subsection{An upper bound for (A)}
We apply Corollary \ref{BS Deco} to the expression denoted by (A). Choosing $A_t=\Gamma^{4}(BS)(t,X_{t},v_{t})$ 
and $B=\frac12\left(\frac{1}{8}D[M,M]_{t}\right)^{2}$, we obtain

\begin{eqnarray*}
(A)&=&\frac{1}{2}\Gamma^{4}(BS)(t,X_{t},v_{t}) \left(\frac{1}{8}D[M,M]_{t}\right)^{2}\\
&+&\frac{1}{1024}\mathbb{E}_{t}\left[\int^{T}_{t} e^{-r(u-t)} \Gamma^{6}(BS)(u,X_{u},v_{u}) \left(D[M,M]_{u}\right)^{2} d[M,M]_{u}\right]
\end{eqnarray*}

In particular, for the Heston model, we obtain

\begin{eqnarray*}
(A)&=&\frac{1}{2}\Gamma^{4}(BS)(t,X_{t},v_{t}) \left(\frac{1}{8}D[M,M]_{t}\right)^{2}\\
&+&\frac{\nu^{6}}{1024}\mathbb{E}_{t}\left[\int^{T}_{t} e^{-r(u-t)} \Gamma^{6}(BS)(u,X_{u},v_{u}) \left(\int_{u}^{T}E_u\left(\sigma _s^2\right)
\varphi(s)^2ds\right)^{2} \sigma^{2}_{u}\varphi^{2}(u)du\right]\\
\end{eqnarray*}
Next, using Lemma \ref{fitage}, setting $a_{u}:=v_{u}\sqrt{T-u}$, and using the fact that $\varphi(t)$ is a decreasing function, we obtain

\begin{eqnarray*}
&&\left|(A)-\frac{1}{2}\Gamma^{4}(BS)(t,X_{t},v_{t}) \left(\frac{1}{8}D[M,M]_{t}\right)^{2}\right|\\
&\leq&C\frac{\nu^{6}}{1024}\mathbb{E}_{t}\left[\int^{T}_{t} e^{-r(u-t)} \left(\frac{1}{a^{7}_{u}}+\frac{5}{a^{6}_{u}}+\frac{10}{a^{5}_{u}}+\frac{10}{a^{4}_{u}}+\frac{5}{a^{3}_{u}}+\frac{1}{a^{2}_{u}}\right) \sigma^{2}_{u}\varphi(u)^{4}du\right]\\
\end{eqnarray*}
It follows from Lemma \ref{tercerlema} (ii) that $\sigma^{2}_{t}\leq \frac{a^{2}_{t}}{\varphi(t)}$ and 

\begin{eqnarray*}
&&\left|(A)-\frac{1}{2}\Gamma^{4}(BS)(t,X_{t},v_{t}) \left(\frac{1}{8}D[M,M]_{t}\right)^{2}\right|\\
&\leq&C\frac{\nu^{6}}{1024}\int^{T}_{t} e^{-r(u-t)} \left(\frac{1}{a^{5}_{u}}+\frac{5}{a^{4}_{u}}+\frac{10}{a^{3}_{u}}+\frac{10}{a^{2}_{u}}+\frac{5}{a_{u}}+1\right) \varphi(u)^{3}du.
\end{eqnarray*}
Now, Lemma \ref{tercerlema} (i) implies that $a_{t}\geq\frac{\sqrt{\theta \kappa}}{\sqrt{2}}\varphi(t)$. Applying the estimate
$\varphi(t)\leq \frac{1}{\kappa}$, we obtain
\begin{eqnarray*}
&&\left|(A)-\frac{1}{2}\Gamma^{4}(BS)(t,X_{t},v_{t}) \left(\frac{1}{8}D[M,M]_{t}\right)^{2}\right|\\
&\leq&C\frac{\nu^{6}}{1024 \kappa^{3}}\left(\frac{4\sqrt{2}\kappa^{2}\sqrt{\kappa}}{\theta^{2}\sqrt{\theta}}+\frac{10\kappa^{2}}{\theta^{2}}+\frac{20\sqrt{2}\kappa\sqrt{\kappa}}{\theta\sqrt{\theta}}+\frac{20\kappa}{\theta}+\frac{5\sqrt{2}\kappa}{\sqrt{\theta\kappa}}+1\right)\int^{T}_{t} e^{-r(u-t)}  du.
\end{eqnarray*}

\subsection{An upper bound for (B)}
We apply the Corollary \ref{BS Deco} to the expression denoted by (B). Choosing $A_t=\Gamma^{3}(BS)(t,X_{t},v_{t})$ 
and $B=\frac{1}{16}D[M,D[M,M]]_{t}$, we get 
\begin{eqnarray*}
(B)&=&\frac{1}{16} \Gamma^{3}(BS)(t,X_{t},v_{t}) D[M,D[M,M]]_{t}\\
&+&\frac{1}{128}\mathbb{E}_{t}\left[\int^{T}_{t} e^{-r(u-t)} \Gamma^{5}(BS)(u,X_{u},v_{u}) D[M,D[M,M]]_{u} d[M,M]_{u}\right]\\
&+&\frac{1}{24}\mathbb{E}_{t}\left[\int^{T}_{t} e^{-r(u-t)} \Gamma^{4}(BS)(u,X_{u},v_{u}) d[M,D[M,D[M,M]]]_{u}\right].
\end{eqnarray*}
In the case of the Heston model, setting $a_{u}:=v_{u}\sqrt{T-u}$ and using Lemma \ref{fitage} and the fact that $\varphi(t)$ is a decreasing function, we obtain
\begin{eqnarray*}
&&\left|(B)-\frac{1}{16} \Gamma^{3}(BS)(t,X_{t},v_{t}) D[M,D[M,M]]_{t}\right|\\
&\leq&C\frac{\nu^6}{128}\mathbb{E}_{t}\left[\int^{T}_{t} e^{-r(u-t)} \left(\frac{1}{a^{7}_{u}}+\frac{4}{a^6_{u}}+\frac{6}{a^{5}_{u}}+\frac{4}{a^{4}_{u}}+\frac{1}{a^{3}_{u}}\right)  \sigma^{2}_{u}\varphi(u)^{6} du\right]\\
&+&C\frac{\nu^6}{24}\mathbb{E}_{t}\left[\int^{T}_{t} e^{-r(u-t)} \left(\frac{1}{a^{7}_{u}}+\frac{3}{a^{6}_{u}}+\frac{3}{a^{5}_{u}}+\frac{1}{a^{4}_{u}}\right)  \sigma^{2}_{u}\varphi^{6}(u) du\right].
\end{eqnarray*}
Now, Lemma \ref{tercerlema} (ii) implies that $\sigma^{2}_{t}\leq \frac{a^{2}_{t}}{\varphi(t)}$ and 

\begin{eqnarray*}
&&\left|(B)-\frac{1}{16} \Gamma^{3}(BS)(t,X_{t},v_{t}) D[M,D[M,M]]_{t}\right|\\
&\leq&C\frac{\nu^6}{128}\mathbb{E}_{t}\left[\int^{T}_{t} e^{-r(u-t)} \left(\frac{1}{a^{5}_{u}}+\frac{4}{a^4_{u}}+\frac{6}{a^{3}_{u}}+\frac{2}{a^{4}_{u}}+\frac{1}{a_{u}}\right)  \varphi(u)^{5} du\right]\\
&+&C\frac{\nu^6}{24}\mathbb{E}_{t}\left[\int^{T}_{t} e^{-r(u-t)} \left(\frac{1}{a^{5}_{u}}+\frac{3}{a^{4}_{u}}+\frac{3}{a^{3}_{u}}+\frac{1}{a^{2}_{u}}\right)  \varphi^{5}(u) du\right].
\end{eqnarray*}
It follows from Lemma \ref{tercerlema} (i) that $a_{t}\geq\frac{\sqrt{\theta \kappa}}{\sqrt{2}}\varphi(t)$. In addition, the inequality 
$\varphi(t)\leq \frac{1}{\kappa}$ implies that  

\begin{eqnarray*}
&&\left|(B)-\frac{1}{16} \Gamma^{3}(BS)(t,X_{t},v_{t}) D[M,D[M,M]]_{t}\right|\\
&\leq&C\frac{\nu^6}{128}\left(\frac{4\sqrt{2}}{\theta^{2}\kappa^{2}\sqrt{\theta\kappa}}+\frac{16}{\theta^{2}\kappa^{3}}+\frac{12\sqrt{2}}{\theta\kappa^{3}\sqrt{\theta\kappa}}+\frac{4}{\theta\kappa^{4}}+\frac{\sqrt{2}}{\kappa^{4}\sqrt{\theta\kappa}}\right)\int^{T}_{t} e^{-r(u-t)}  du\\
&+&C\frac{\nu^6}{24}\left(\frac{4\sqrt{2}}{\theta^{2}\kappa^{2}\sqrt{\theta\kappa}}+\frac{12}{\theta^{2}\kappa^{3}}+\frac{6\sqrt{2}}{\theta\kappa^{3}\sqrt{\theta\kappa}}+\frac{2}{\theta\kappa^{4}}\right)\int^{T}_{t} e^{-r(u-t)} du.
\end{eqnarray*}
\end{appendices}
 
\end{document}